\documentclass[12pt,lettersize]{emulateapj}
\usepackage{apjfonts} 

\newcommand{\etal}{et~al.}
\newcommand{\eg}{e.g.,}

\newcommand{\ie}{i.e.,}
\newcommand{\spitzer}{{\it Spitzer}}

\newcommand{\zmed}{z_{\mbox{\small{med}}}}
\newcommand{\mum}{$\mu$m}
\newcommand{\muj}{$\mu$Jy}
\newcommand{\OII}{[O\,II]}

\def\spose#1{\hbox to 0pt{#1\hss}}
\def\simlt{\mathrel{\spose{\lower 3pt\hbox{$\mathchar"218$}}
     \raise 2.0pt\hbox{$\mathchar"13C$}}}
\def\simgt{\mathrel{\spose{\lower 3pt\hbox{$\mathchar"218$}}
     \raise 2.0pt\hbox{$\mathchar"13E$}}}

\shorttitle{Photometric Redshifts in the IRAC Shallow Survey}
\shortauthors{Brodwin et al.}

\begin{document}


\title{Photometric Redshifts in the IRAC Shallow Survey}


\author{M.~Brodwin\altaffilmark{1}, 
M.~J.~I.~Brown\altaffilmark{2}, 
M.~L.~N.Ashby\altaffilmark{3}, 
C.~Bian\altaffilmark{4},
K.~Brand\altaffilmark{5}, 
A.~Dey\altaffilmark{5}, 
P.~R.~Eisenhardt\altaffilmark{1}, 
D.~J.~Eisenstein\altaffilmark{6},
A.~H.~Gonzalez\altaffilmark{7},
J.--S.~Huang\altaffilmark{3}, 
B.~T.~Jannuzi\altaffilmark{5}, 
C.~S.~Kochanek\altaffilmark{8},
E.~McKenzie\altaffilmark{7},
M.~A.~Pahre\altaffilmark{3}, 
H.~A.~Smith\altaffilmark{3}, 
B.~T.~Soifer\altaffilmark{4}
S.~A.~Stanford\altaffilmark{9,10},
D.~Stern\altaffilmark{1},
\& R.~J.~Elston\altaffilmark{7,11}}


\altaffiltext{1}{Jet Propulsion Laboratory, California Institute of
  Technology, Mail Stop 169-506, Pasadena, CA 91109}
\altaffiltext{2}{Princeton University Observatory, Peyton Hall, Princeton University, Princeton, NJ 08544}
\altaffiltext{3}{Harvard--Smithsonian Center for Astrophysics, 60 Garden Street, Cambridge, MA 02138}
\altaffiltext{4}{\spitzer Science Center, MC 220--6, California Institute of Technology, 1200 East California Boulevard, Pasadena, CA 91125}
\altaffiltext{5}{National Optical Astronomy Observatory, 950 N.~Cherry Ave., Tucson, AZ 85719}
\altaffiltext{6}{Steward Observatory, University of Arizona, 933 N.~Cherry Ave., Tucson, AX 85121}
\altaffiltext{7}{Department of Astronomy, University of Florida, Gainesville, FL 32611}
\altaffiltext{8}{Department of Astronomy, The Ohio State University, 140 West 18th Avenue, Columbus, OH 43210}
\altaffiltext{9}{University of California, Davis, CA 95616}
\altaffiltext{10}{Institute of Geophysics and Planetary Physics, Lawrence Livermore National Laboratory, Livermore, CA 94550}
\altaffiltext{11}{Deceased}


\begin{abstract}

Accurate photometric redshifts are calculated for nearly
200,000 galaxies to a 4.5 micron flux limit of $\sim13$
\muj\ in the 8.5 deg$^2$ \spitzer/IRAC Shallow survey.  Using a
hybrid photometric redshift algorithm incorporating both
neural--net and template--fitting techniques, calibrated with
over 15,000 spectroscopic redshifts, a redshift accuracy of
$\sigma = 0.06\,(1+z)$ is achieved for 95\% of galaxies at
$0<z<1.5$.  The accuracy is $\sigma = 0.12\,(1+z)$ for 95\% of
AGN at $0 < z < 3$.  Redshift probability functions, central to
several ongoing studies of the galaxy population, are computed
for the full sample.  We demonstrate that these functions
accurately represent the true redshift probability density,
allowing the calculation of valid confidence intervals for all
objects.  These probability functions have already been used to
successfully identify a population of \spitzer--selected high
redshift ($z>1$) galaxy clusters.  We present one such
spectroscopically confirmed cluster at $\left<z\right>=1.24$,
ISCS J1434.5+3427.  Finally, we present a measurement of the
4.5\mum--selected galaxy redshift distribution.

\end{abstract}





\section{Introduction}

In modern wide-field imaging surveys, accurate photometric redshifts
have become an indispensable tool for studying the evolving galaxy
population.  This approach allows studies of the galaxy luminosity and
correlation functions, power spectrum analyses, and rare--object
searches, reserving costly 8--m class spectroscopic follow--up for
analyses where such data are crucial.

Much of the new extragalactic science enabled by the {\it Spitzer
  Space Telescope} will not just be more efficient with photometric
redshifts, but rather completely reliant on them.  \spitzer\ is
uniquely capable of detecting huge numbers of galaxies from their rest
1.6 \mum\ stellar emission or mid--IR PAH features all the way to
$z=2$ and beyond.  These objects are often quite faint in the optical,
due to either quiescence or heavy extinction, making spectroscopic
follow--up difficult or impossible with existing facilities. 

The IRAC Shallow survey \citep{eisenhardt04} is a wide--field 8.5
deg$^2$ \spitzer/IRAC imaging survey in the NOAO Deep Wide--Field
Survey \citep[NDWFS;][]{ndwfs99} Bo\"otes field, designed to study
galaxy formation and evolution across a wide range of redshifts, mass
scales, colors and environments.  Extensive complementary data
includes a 5 ksec per pointing X-ray mosaic (XBo\"otes;
\citealt{xbootes1}; \citealt{xbootes2}; \citealt{xbootes3}), and
\spitzer/MIPS 24/70/160 \mum\ imaging (GTO program, Soifer PI) over
the full survey area.  In addition, the FLAMINGOS Extragalactic Survey
\citep[FLAMEX;][]{elston06} provides deep $J$ and $K_s$ imaging over
the central $\approx 4.7$ deg$^2$.  Finally, there are 17,017
spectroscopic redshifts of galaxies to $z\sim 0.8$ and AGN to $z =
5.85$ from the AGN and Galaxy Evolution Survey (AGES;
\citealt{cool06}; Kochanek \etal\ in preparation), along with $\sim
500$ redshifts of fainter, higher redshift galaxies from various other
surveys in Bo\"otes, which are available for calibration of the
photometric redshift algorithm.

In this paper, we combine the optical photometry from the NDWFS, the
near--IR photometry from FLAMEX, and the mid--IR photometry from the
IRAC Shallow survey to compute accurate photometric redshifts for
4.5\mum--selected galaxies to $z=1.5$ and AGN to $z=3$.  A hybrid
method, combining a standard template--fitting algorithm with an
artificial neural net--based approach, is developed to make optimal
use of the large AGES spectroscopic calibration sample.

Redshift probability functions are derived for the full sample, nearly
200,000 objects over 8.5 deg$^2$, and are employed to calculate the
4.5\mum\ redshift distribution.  They are also the input to a
high--redshift cluster search underway in Bo\"otes, which has already
yielded the highest redshift galaxy cluster to date
\citep[$\left<z\right>=1.41$;][]{stanford05}.  This paper presents a
new, spectroscopically confirmed high redshift cluster at
$\left<z\right>=1.24$.

This paper is organized as follows.  In \textsection{\ref{Sec: data}}
the various data sets used in this paper are briefly described.  In
\textsection{\ref{Sec: Method}} the photometric redshift algorithm is
presented, and the results are compared to the AGES spectroscopy.  The
calculation of redshift probability functions is described in
\textsection{\ref{Sec: RPF}}, and in \textsection{\ref{Sec:
    Applications}} we employ them to compute the 4.5\mum\ redshift
distribution and present a new $\left<z\right>=1.24$ galaxy cluster
discovered using them.  We summarize our results in
\textsection{\ref{Sec: Summary}}.  All magnitudes are Vega--based.

\section{Photometric and Spectroscopic Data}
\label{Sec: data}

\subsection{Optical and NIR Photometry}
Optical $B_W$, $R$, and $I$ imaging data were taken from the third
data release (DR3) of the public NDWFS survey in the Bo\"otes field.
These data, obtained with the Mosaic--I camera on the NOAO 4--m, are
described fully in Jannuzi \etal\ (in preparation) and available
through the NOAO Science Archive (http://archive.noao.edu/nsa/).

Robust photometric errors were estimated via Monte--Carlo simulation,
and extensive flagging of both pixel and image artifacts allows
selection of high-quality photometric samples over the full survey
area.  The optical photometry reaches 3$\sigma$ 5\arcsec\ diameter
Vega depths of $B_W=25.3$, $R= 24.1$, and $I=23.6$.  The large
apertures are taken to better match the IRAC PSF and to minimize the
effect of seeing variations in the optical data, as discussed below.
The NDWFS is significantly deeper for point source or small aperture
photometry.

The FLAMEX survey \citep{elston06} is a near--IR $J$-- and $K_s$--band
imaging survey undertaken with the FLAMINGOS camera on the NOAO
2.1--m.  Photometric errors were determined via extensive Monte Carlo
simulations, accounting for PSF variations across the field.  The
survey covers the central $\sim$4.7 deg$^2$ subset of the Bo\"otes
field to a 5\arcsec\ diameter aperture, 50\% completeness limit of
$K_s= 19.4$.

\subsection{IRAC imaging}

The IRAC Shallow survey, introduced in \citet{eisenhardt04}, is a
\spitzer/IRAC imaging survey in the NDWFS Bo\"otes field, covering 8.5
deg$^2$ with 3 or more 30 second exposures per position at 3.6, 4.5,
5.8, and 8.0 \mum\ to 5$\sigma$ flux limits in a 5\arcsec\ aperture of
10.0, 13.3, 78.0, and 68.3 \muj, or Vega magnitude limits of 18.6,
17.8, 15.4, and 14.9, respectively.

Separate photometric catalogues were extracted in each channel using
SExtractor $2.3.2$ \citep{sextractor} in double--image mode, producing
matched catalogs in the other 3 IRAC bands.  The detection images in
each channel were weighted by the error images generated by the MOPEX
mosaicking software \citep{mopex}.  This paper focusses on the 4.5
\mum--selected catalog, which is the natural selection band for the
$1< z< 2$ cluster search described below (\eg\ Eisenhardt \etal, in
preparation).

Quality control was maintained through the extensive use of flags.  In
particular, to ensure reliable colors, objects with less than 3
exposures in any single pixel within the aperture of interest were
rejected from the final catalogs.  This primarily removes objects from
the edges of the field, with only $\sim$1\% of the 4.5 \mum\ sample
rejected in the main overlap region.  This spatial selection function
is well quantified and will not affect any science analyses.  This
flagging was carried out for all of our apertures, which span
diameters of 1--20\arcsec, chosen to match the DR3 NDWFS catalogs that
form our primary complementary data set.  All IRAC aperture photometry
was corrected to large (24.4\arcsec\ diameter) apertures to account
for PSF losses.  For the 5$\sigma$ 5\arcsec\ diameter aperture of
interest in this work, there are 211, 260 objects which have the full
exposure time in both the [3.6] and [4.5] bands.

\subsubsection{Catalog Matching}

There is a small offset, of 0.38\arcsec\ in right ascension and
0.15\arcsec\ in declination, between the astrometric solutions of the
near-- and mid--IR catalogs, which are tied to the 2MASS reference
frame, and the optical catalog, which is tied to the USNO--A2
\citep{monet03}.  This offset, caused in part by errors in centroiding
the bright Tycho--II stars used to zeropoint the USNO--A2 astrometry,
was removed prior to matching (for further details, see Jannuzi
\etal, in preparation).

Detections in the optical and near--IR were matched to the 4.5
\mum\ sources if the centroids were within $1\arcsec$ of each other.
For extended objects, detections in the different bands were matched
if the centroids were within an ellipse defined using the second order
moments of the light distribution of the object\footnote{This ellipse
  was defined with the SExtractor parameters $2 \times {\rm
    A\_WORLD}$, $2 \times {\rm B\_WORLD}$, and ${\rm THETA\_WORLD}$.}.
IRAC--selected objects with no match in a given optical or near--IR
band were assigned a Monte Carlo estimated 1$\sigma$ flux limit
representing the sky variation in a 5\arcsec\ diameter aperture.

\subsubsection{Photometric Redshift Sample}
\label{Sec: Photo-z Sample}

We define the photometric redshift sample as the subset of the
[4.5]--selected matched catalog for which both the [3.6] and [4.5]
data have the full 90s exposure time, and for which at least 2 of the
3 optical bands contain useful (i.e.,~unmasked) photometric data.
Note that object {\it detections} in the optical, near--IR, and [3.6]
bands are not required; 1$\sigma$ limits are used for non--detections.
This results in a final multi--wavelength photometric redshift sample
of 194,466 objects.

\subsection{Spectroscopic Redshifts}

The large sample of spectroscopic redshifts from AGES (Kochanek
\etal\ in preparation) provides a crucial training sample for the
photometric redshift algorithm.  AGES is a wide--field MMT/Hectospec
\citep{hectospec} redshift survey, version 2.0 of which contains high
quality spectroscopic redshifts for 17,017 objects, including galaxies
to $z \sim 0.8$ and AGN to $z = 5.85$ \citep{cool06}.  From this
sample, 15,052 objects correspond to sources in the photometric
redshift sample defined above.

The AGES survey was designed to allow magnitude--limited samples to be
selected as a function of wavelength from the X--ray to the radio.  In
the IRAC 4.5\mum\ band, the AGES survey is $\approx100$\% complete to 15.2 mag
(150 \muj), and statistically complete (with 30\% random sampling) to
15.7 mag (95 \muj).

However, more than half the spectroscopic sample lies beyond the
fainter of these limits, driven primarily by the optical magnitude
limits of $I =20$ for extended objects and $I=21.5$ for point sources,
where the latter limit was designed to include relatively large
numbers of AGN and quasars.  In addition, the complete samples in the
X-ray, UV, Far--IR and radio result in the survey being overweight in
both active galaxies and those which are strongly starforming.

We have also assembled a deep ($R\sim 25$) heterogeneous sample of
spectroscopic redshifts from several ongoing projects in the Bo\"otes
field for the purposes of calibrating the photometric redshifts.
These projects include spectroscopic studies of optical, near-IR, IRAC
and MIPS--selected sources, including our own spectroscopy of $1.1 < z
< 1.4$ galaxy cluster members described below.  As many of the
principal investigators of these projects are also co--investigators
of the present work (AD, BTJ, BTS, DS, SAS), we will call this the
``in--house'' sample for brevity.

\newpage 
\section{Method:  Photometric Redshifts}
\label{Sec: Method}
The impressive accuracy of photometric redshift algorithms of various
types
\citep{fernandez-soto02,benitez,fontana00,sawicki97,connolly97,firth02,brodwin06,rcs_photo-z,combo17_LF}
is a good indication that the field is rapidly maturing at least as
concerns optical/NIR surveys.  It is less clear which methodology is
best for redshift estimation from photometry extending into the $L$
and $M$ bands and beyond.  A host of physics not included in current
population synthesis models, including ubiquitous PAH emission and
molecular and silicate absorption, could potentially complicate
photometric redshift estimation in the first generation of
\spitzer\ surveys.

Well--calibrated empirical methods such as fitting functions
\citep[e.g.][]{connolly95,brunner00} and neural nets
\citep{firth03,ANNz,vanzella04} are capable of matching or surpassing
the accuracy of template--fitting methods.  They can be particularly
powerful for analyses in which it is sufficient to predict accurate
redshifts for a relatively small subset of the general galaxy
population, albeit one with a large, representative spectroscopic
training set.  For instance, \citet{blake06} and \citet{padmanabhan06}
measured the galaxy power spectrum using only bright, low-redshift,
early type galaxies, selected with a variety of color and magnitude
cuts.  The large 4000 \AA\ breaks and lack of SED ambiguity produces
highly accurate redshifts in these samples.

In this paper we wish to develop a methodology with optimal
photometric redshift accuracy over a much larger redshift baseline and
spanning {\em all} galaxy types.  Not surprisingly there is no
spectroscopic sample spanning the full range of galaxy color,
magnitude, spectral type, and redshift in the present survey.  Despite
the large size of the AGES spectroscopic sample, it consists of the
low redshift, high luminosity component of the general IRAC
photometric redshift sample.  The situation for QSOs and AGN is more
encouraging.  The broad range of targeting methods and deeper limiting
magnitudes likely lead to a broadly representative sample of
unobscured AGN for all redshifts.

On the other hand, template--fitting methods calibrated with even
modest spectroscopic redshift samples produce photometric redshifts
which are generally robust outside of the narrow parameter space in
which they are specifically validated.  For this reason
template--fitting algorithms are the de facto standard in the
literature.  An important advantage of this technique is the
straightforward generation of redshift probability functions from the
likelihood analysis, which are key to many science applications.  This
method is only accurate if the galaxy templates are representative of
the observed galaxies.  Strong PAH-emitting galaxies, for which
reliable templates do not yet exist, along with quasars and AGN which
have minimal continuum breaks, present a challenge for this technique.

In the IRAC Shallow survey we therefore employ a hybrid approach in
which a template--fitting algorithm is used as the core method,
supplemented by a well-calibrated neural net technique for the small
subset of objects which both require and merit it.

\subsection{Empirical Template--Fitting Algorithm}

The template--fitting algorithm closely follows that described in
detail in \citet{brodwin06}.  Coleman, Wu \& Weedman (1980, CWW)
galaxy SEDs, supplemented by the \citet{kinney96} empirical starburst
(SB3 and SB2) SEDs are used as basis templates.  These templates were
extended to the far--UV and mid--IR using \citet{bc03} models.  To
improve the redshift accuracy linear interpolates were derived,
resulting in 19 templates finely spanning the template space between
the CWW Elliptical and SB2.

Population synthesis codes do not yet accurately model the complicated
physical processes, in particular strong PAH emission, which can
dominate the rest--frame $\lambda \ga 5$ \mum\ emission from normal
galaxies. In view of this, and given that only the brightest $\sim
10\%$ of our sample has well--measured fluxes in the $5.8$ and $8.0$
\mum\ bands, we elected to limit the fitting to the $\lambda < 5$
\mum\ regime where the models are expected to better approximate the
true SEDs.  This mild restriction still permits us to sample the
stellar peak at 1.6 \mum, a useful redshift indicator (\eg\
\citealt{simpson99}; \citealt{sawicki02}) out to $z\sim2$.  The PAH
emission at $\lambda \sim 3.3$ \mum, which could potentially cause a
template mismatch for strongly starforming galaxies, is quite modest
in terms of equivalent width \citep[EW $= 0.02 $\mum;][]{lu03}.
Indeed, this feature has only 0.5\% of the power of the PAH features
longward of 5\mum\ \citep{helou00}.  Since it redshifts out of the
$[4.5]$ band by $z\sim 0.5$, any deleterious effect on redshift
estimation should be limited to low redshifts.

Redshifts were fitted between $0 \le z \le 5$ using 5\arcsec\ diameter
aperture $B_WRI+JK_s+[3.6][4.5]$ photometry where available.  Reliable
detections are not required as this would impose a strong selection
effect; limits are used for areas of sky which were observed but for
which no object was detected.  Many deeper IRAC surveys suffer from
considerable confusion in the bluer bands, leading to substantial
difficulties in deriving accurate photometry.  Due to the combination
of area and depth targeted in the Shallow survey, the images are not
confused, allowing straightforward aperture photometry.

To obtain robust galaxy colors, it is common in ground-based imaging
surveys to smooth the images in all bands to a common worst seeing.
Due to the relatively large PSF mismatch between
\spitzer\ ($\sim$2\arcsec) and the ground--based optical data
(0.8\arcsec-1.3\arcsec), it is not clear that this is the best
approach.  The large apertures required to enclose a substantial
fraction of infrared light, empirically 5\arcsec, are sufficient to
minimize the effects of seeing variations in the optical and near--IR
images.  We have verified that photometric redshifts computed with
smaller apertures (3\arcsec) exhibit redshift-- and
position--dependent systematic errors, likely due to ground--based
seeing variations.  These effects vanish with the larger
5\arcsec\ aperture photometry.

\subsection{Correcting Templates and Zero Points}

Comparison of preliminary photometric redshifts with AGES spectroscopy
indicated an error in the mid--IR color of model elliptical galaxies.
The spectral synthesis codes at $\lambda > 1$ \mum\ model a
zero--color Rayleigh--Jeans tail typical of simple stellar
populations.  Recent \spitzer\ studies of nearby galaxies
\citep{pahre04} indicate that, while reasonable for late type
galaxies, early--type galaxies are in fact blue in the mid--IR, with
$[3.6]-[4.5] \approx -0.15$, due to CO absorption.  To account for
this the elliptical template (in $F_{\lambda}$) was scaled down
between $1 < \lambda (\mu \mbox{m}) < 5$ by a factor of $m\,(\lambda
(\mu \mbox{m})-1)$, where the slope, $m = -0.11$, was determined by
maximizing the photometric redshift accuracy for high S/N AGES
elliptical galaxies. This change was carried through the template
interpolations.

The large AGES spectroscopic sample was also used extensively to
analyze the absolute inter--survey photometric calibration.  For those
objects well characterized by the model templates, the secure
spectroscopic redshifts allow the AGES sample to be effectively used
as spectrophotometric standards to determine inter--survey photometric
offsets.  These were found to be negligible in the optical and small
in the near--IR, $\Delta J=-0.10$ mag and $\Delta K_s = -0.02$ mag.
However, in the mid-IR the offsets between the observed photometry and
the \citet{bc03} models were considerably larger, $\Delta[3.6] = 0.27$
mag and $\Delta[4.5] = 0.32$ mag.

While systematic zeropoint errors at the $\sim 5-10\%$ level are
possible with IRAC data \citep{reach05}, this much larger error is
likely due to the inadequacy of the spectral synthesis models in the
near-- and mid--IR.  Indeed, as illustrated recently by
\citet{maraston05}, the near-- and mid--IR to optical colors predicted
by independent spectral synthesis models
\citep[\eg][]{bc03,pegase2,starburst99,vazdekis96} have a scatter of
0.2--0.3 mag even for identical input stellar evolutionary tracks.  In
addition, the new population synthesis models by \citet{maraston05},
which include the contributions from the post main sequence
evolutionary phases, predict higher infrared fluxes than previous
models, which are more consistent with our corrected aperture
photometry.  Work is in progress (Kochanek \etal\ in preparation) to
empirically derive low resolution rest--frame mid--IR SEDs for both
galaxies and AGN using the AGES spectroscopic sample and the full
multi--wavelength photometry in the Bo\"otes field.  For the present
paper we adopt the above magnitude offsets as empirically motivated
corrections to the photometry--SED combination that optimize
photometric redshift accuracy.

This calibrated template--fitting algorithm was tested on AGES objects
with $z\la1$, as all $z>1$ AGES objects are QSOs or AGN and of little
use in evaluating the performance of galaxy template--fitting
algorithms.  As shown in the top panel of Figure \ref{Fig:
  just_templates}, the results are quite poor, with an rms dispersion
in the photometric redshifts about the true redshifts of $\sigma
\approx 0.38$, or $\sigma \approx (1+z)\,0.27$.  The dispersion for a
95\% clipped sample, the 95\% of objects with the smallest absolute
redshift difference, is significantly better, with $\sigma \approx
0.09$, or $\sigma \approx 0.07\,(1+z)$.  While indicative that in the
mean the method works quite well, there are clearly objects not well
fit by the empirical templates.
\begin{figure}[bthp]
\epsscale{1.25}
\hspace*{-0.65cm}
\plotone{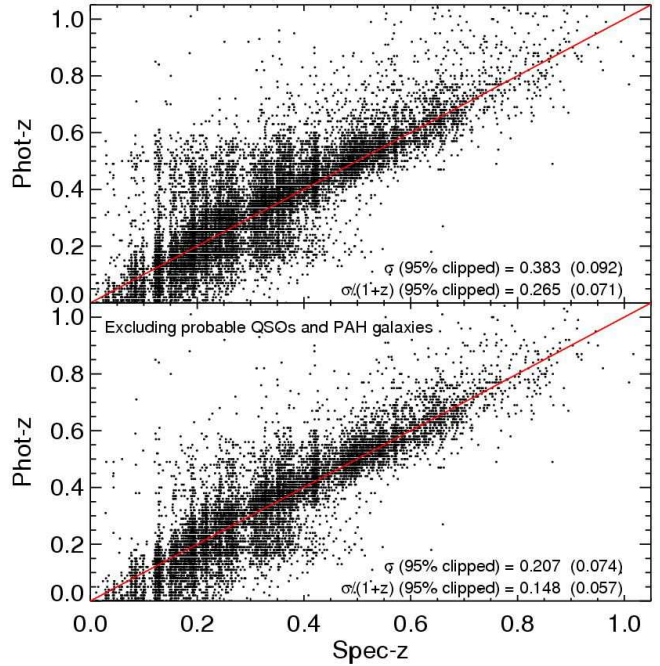}
\caption{Comparison of calibrated template--fitting photometric
  redshifts with AGES spectroscopy.  In the top panel all $z<1.1$ AGES
  objects are plotted, whereas in the bottom panel likely AGN and
  strongly PAH--emitting galaxies have been removed.  The full rms
  dispersion markedly improves in the bottom panel.  The clipped
  statistics (in parentheses) are quite good in both cases, with
  $\sigma \la 0.1$.}
\label{Fig: just_templates}
\end{figure}
As discussed above, PAH--emitting and active galaxies, two object
classes not represented in the empirical galaxy templates, are
over--represented in the AGES sample.  To see how this affects the
statistics, the lower panel of Figure \ref{Fig: just_templates} omits
objects likely to be either PAH--emitters or AGN.  The former are
defined in this work as objects with an IRAC color of $[5.8]-[8.0] >
1$.  The latter were identified on the basis of their infrared colors
in an IRAC color--color diagram similar to the one described in detail
in \citet{stern05}.  That paper demonstrated that active galaxies can
be readily identified based solely on their infrared colors
\citep[also see][]{lacy04}.  Omitting these two populations removes
70\% of the variance, or about half the rms error, while maintaining
similar clipped statistics.  We now turn to techniques of estimating
redshifts for these populations.

\subsection{Artificial Neural Net Algorithm}

Although the utility of purely empirical methods is limited to the
parameter space of the calibration sample, such methods do offer
several unique advantages.  Neither absolute band--to--band
calibrations, nor a complete knowledge of the galaxy population are
required for accurate redshift estimation.  Given a large
spectroscopic training set, algorithms such as polynomial fits or
artificial neural nets (ANNs) can be trained to predict the redshifts
of objects of all types using the actual survey photometry.

Regardless of method, the uniqueness of the mapping from colors to
redshifts is the underlying limitation in redshift accuracy.  Objects
with approximately power--law spectra such as QSOs will never allow
very accurate photometric redshift estimation.  Nevertheless,
empirical methods do in principle allow the best redshift estimation
possible for each population.  For populations with strong spectral
features but lacking accurate template spectra, empirical algorithms
are expected to show marked improvement over template--fitting
methods.

To improve the redshift estimation of $z<0.5$ PAH--emitting starburst
galaxies, as well as very active galaxies, for which the AGN component
dominates the observed mid--IR emission, the AGES spectroscopic sample
was used to train a neural net algorithm.  We adopted the public code
ANNz \citep{ANNz} for this purpose.

Briefly, the ANN is trained to match a set of observational inputs (in
this case galaxy photometry) to a set of known outputs (the
spectroscopic redshifts) by minimizing a cost function.  The form of
the cost function is determined by the architecture of the ANN, which
specifies the number of hidden nodes between the input and output
nodes.  Following \citet{firth03}, we employ a 7:10:10:10:1
architecture, which takes 7 inputs, the $B_WRI[3.6][4.5][5.8][8.0]$
photometry, has three 10--node hidden layers and a single output node,
the photometric redshift \citep[see][for
  details]{ANNz,firth03,lahav96}.  The inclusion of the $[5.8]$ and
$[8.0]$ photometry (or limits) allow the PAH features in starburst
galaxies and the mid--IR excesses in active galaxies to be fitted, or
at least differentiated.  Only objects observed in all of the above
bands were used in the fit; the near--IR data covering only a subset of
the field was not used.  

The AGES sample was divided into training, validation and testing
subsets, containing 7500, 2500, and 5052 objects, respectively.  The
cost function is fitted on the training set and evaluated on the
validation set after each iteration.  When trained, the algorithm can
be run on the independent testing set to determine the accuracy of the
method.  The ``consensus'' median prediction from a committee of ANNs,
each the result of independent training sessions (initialized with
different random seeds), produces the most reliable redshifts and
error estimates \citep[\eg][]{firth03}.  Although the current
implementation of ANNz does not incorporate the photometric errors in
its determination of the photometric redshift, it does use them to
estimate the redshift uncertainties.  A committee of 10 ANNs was
adopted in this work.

\begin{figure}[bthp]
\epsscale{1.25}
\hspace*{-0.65cm}
\plotone{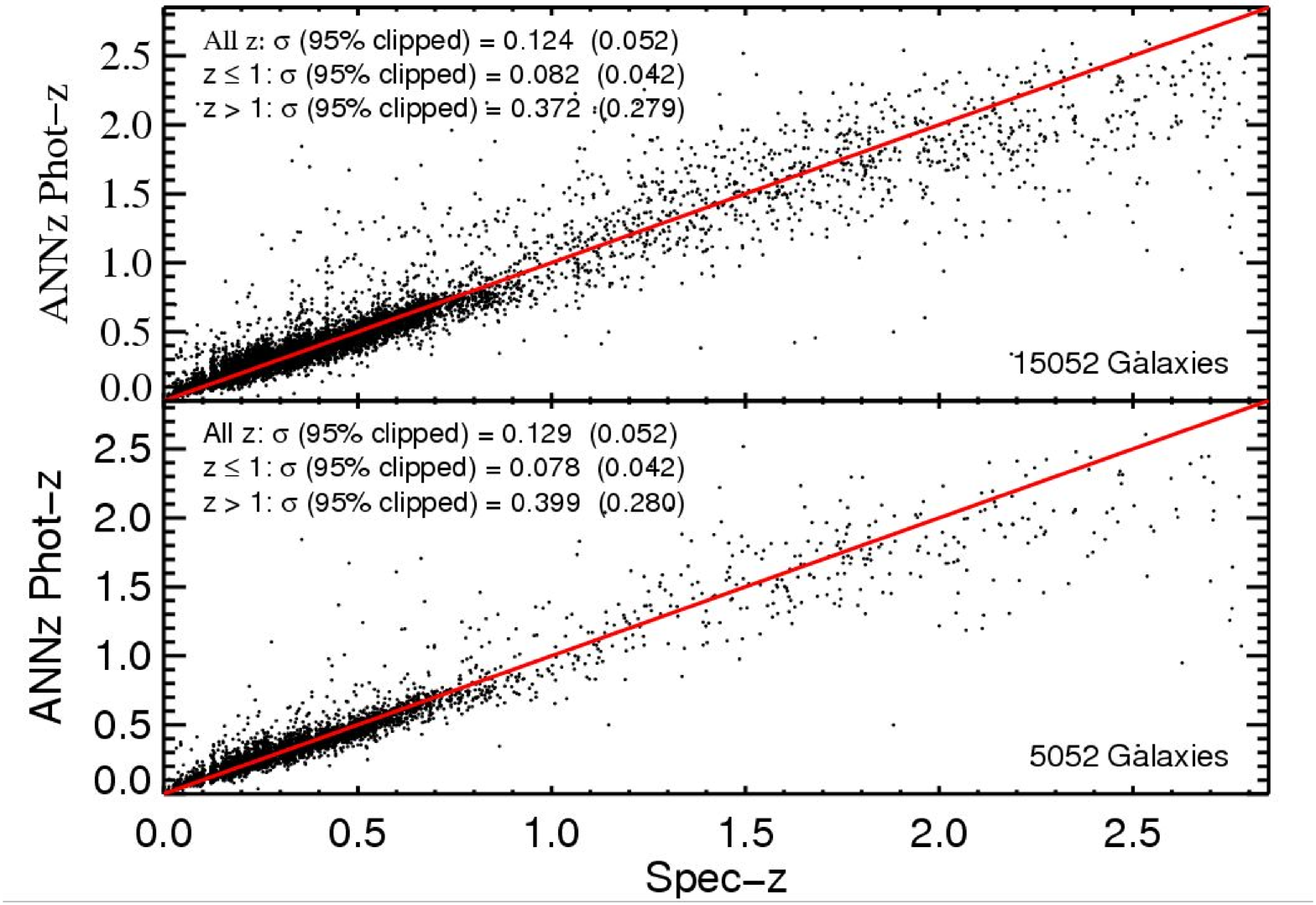}
\caption{ANNz photometric vs.\ spectroscopic redshift for the full
  AGES sample (top panel), and the testing sample alone (bottom).  The
  clipped results are quite similar in both samples, with very
  accurate ($\sigma \sim 0.05$) redshift prediction for the normal
  galaxy sample at $z<1$ (including PAH--emitting galaxies), and more
  modest accuracy ($\sigma \sim0.3-0.4$) at $z>1$ where the sample is
  dominated by QSOs and AGN.}
\label{Fig: ANNz}
\end{figure}

The results are plotted in Figure \ref{Fig: ANNz} for both the full
AGES sample (top panel), and the smaller test set which was
independent of the ANN training.  The results for these two samples
are essentially identical, indicating that our training sample is
large enough to span the distribution of spectral types and redshifts
in the AGES sample.  The dispersion over all redshifts, $\sigma \sim
0.13$, is dominated by the $z>1$ AGN, whose weak broad--band spectral
features lead to greater redshift uncertainty for any algorithm.
Considering just the $z<1$ AGES galaxies, we find the dispersion drops
to $\sigma \sim 0.08$, whereas for the $z>1$ AGN it rises to $\sigma
\sim 0.4$.  This is typical of the accuracy of other attempts to
measure AGN redshifts photometrically
\citep[\eg][]{kitsionas05,babbedge04,weinstein04}.
\subsection{Hybrid Approach}

In assessing the advantages and drawbacks of the two redshift
algorithms presented so far, an obvious complementarity is apparent.
The ANN provides excellent redshift estimation for low--z galaxies
bright enough to have well measured photometry in all 4 IRAC bands,
including strongly PAH--emitting galaxies.  Furthermore, the ANN
redshift accuracy for AGN is superior at all redshifts to that
achievable by template--fitting methods.  Due to the IR--excess of
these active galaxies \citep{stern05}, large numbers of them are
well-measured in all four IRAC bands, at least out to $z\sim2$.
Therefore, within the bright limits of the AGES survey, these two
populations are well represented and hence well calibrated in the ANN.

The template--fitting method, while not reliable for
accretion--dominated active galaxies, is quite robust for normal
galaxies in the $z\la1$ AGES sample, barring low redshift strong
PAH--emitters.  This method has the critical advantage that it is
expected to be robust outside the parameter space in which it was
tested.  In particular it should be reliable to magnitudes much
fainter that the AGES sample, and out to higher redshift.  Another
important advantage of this method is that it allows straightforward
generation of redshift probability functions, which are essential for
many applications.  Finally, it produces restframe properties for the
fitted galaxies, such as absolute magnitudes.

We therefore construct a hybrid photometric redshift sample, combining
the above methods according to their strengths.  While it is tempting
to simply adopt the ANN method for all bright galaxies, we limit its
sphere of influence to the color--selected AGN and PAH--emitters as
defined below.

\begin{figure}[bthp]
\epsscale{1.25}
\hspace*{-0.65cm}
\plotone{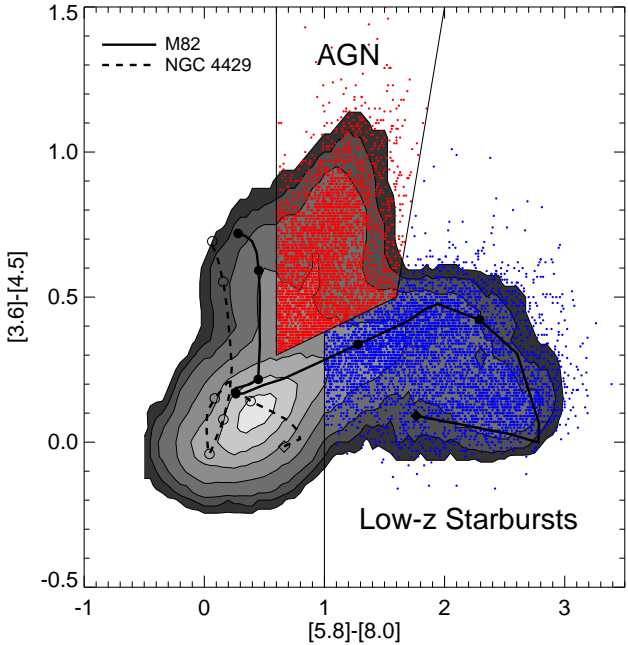}
\caption{IRAC color--color plot showing the color criteria for
  inclusion in the neural net component of the hybrid photometric
  redshift algorithm.  Grey contours illustrate the colors of the full
  photometric redshift sample.  The likely AGN and PAH-emitters are
  overplotted in red and blue points, respectively.  In addition to
  these color selections, objects are subject to strict AGES--defined
  flux limits, as defined in the text.  Model tracks of two $0\le z
  \le 2$ non--evolving galaxy templates from \citet{devriendt99} are
  also plotted.  The solid curve represents M82, a starburst galaxy;
  the dashed curve shows NGC 4429, an S0/Sa galaxy with a star
  formation rate approximately 4000 times lower.  The tracks begin at
  $z=0$ at the diamond symbols, and circular annotations are made at z
  = 0.25, 0.5, 0.75, 1.0, 1.5, and 2.0.}
\label{Fig: wedge}
\end{figure}

Selection of objects for ANN redshift estimation was made using color
cuts in $[3.6]-[4.5]$ vs.\ $[5.8]-[8.0]$ color--color space,
illustrated in Figure \ref{Fig: wedge}.  Model tracks illustrate where
representative quiescent and starburst galaxies appear from $0\le z\le
2$ in this plot.  Objects lying in the ``AGN wedge'', defined in
\citet{stern05}, were classified as likely active galaxies.  On the
other hand, objects outside the AGN wedge and with $[5.8]-[8.0] > 1$
were taken to be potential PAH galaxies.  For both populations we only
estimated redshifts for objects matching the IRAC flux limits of the
full spectroscopic sample, which were $[17.5, 17.0, 16.2, 15.5]$ for
AGN and $[16.75, 16.6, 16.2, 14.25]$ for PAH--emitters in the 3.6,
4.5, 5.8, and 8.0 micron bands, respectively.  The limits are
generally deeper for the active galaxies since unresolved AGN
candidates were targeted to a greater depth in AGES.  No extrapolation
to fainter magnitudes was allowed for either object class.  In
addition, adequate photometric coverage was required in all 7 bands
that were used in the calibration of the ANN.
The strong flux constraints, in particular in the $[5.8]$ and $[8.0]$
bands, limit the sample to the very brightest AGN at $z \la 2.5$ and
PAH--emitters at $z\la 0.5$, as in the AGES sample.  The final
selection included 3681 AGN and 4766 PAH--emitters for which the
template--fitting redshifts were replaced with ANN redshifts.
Representing only $\sim 4\%$ of the galaxy sample, this approach
serves primarily to reduce the number of outliers.

\subsection{Comparison to Spectroscopy}
\label{Sec: Results}

In this Section we demonstrate that the hybrid method surpasses the
simple template--fitting algorithm in redshift accuracy for both
galaxies and AGN, though the improvement is more substantial for the
latter.  We adopt the QSO/AGN targeting criteria employed in the AGES
survey to distinguish between normal and active galaxies.  These
criteria, described fully in Kochanek \etal\ (in preparation),
identify AGN by combining optical morphology with flux or color cuts
in x--ray through radio wavelengths.
\begin{figure*}[bthp]
\plotone{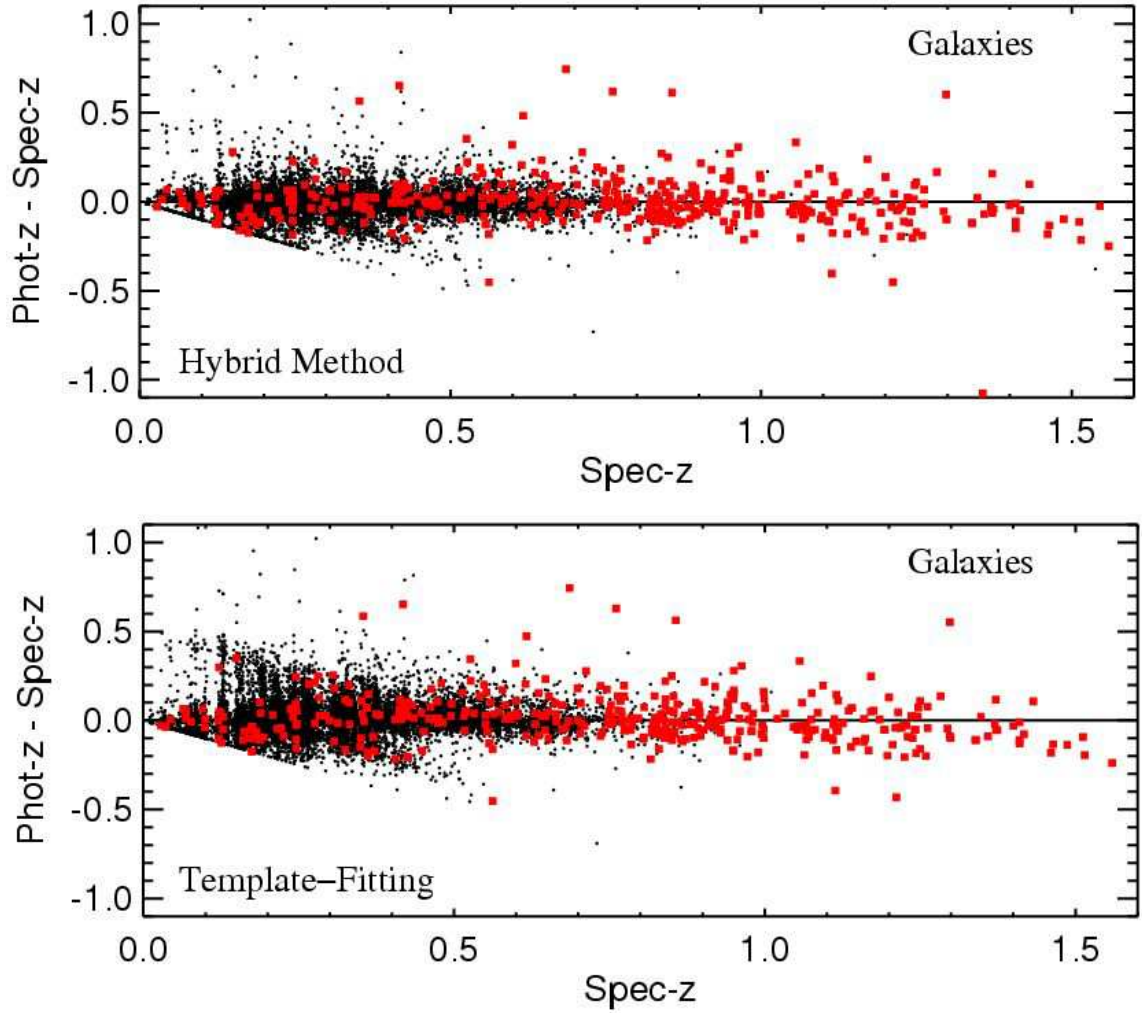}
\caption{Hybrid (top) and template--fitting (bottom) photometric
  redshifts for the AGES galaxy sample (dots) and the deeper,
  in--house spectroscopic sample (filled squares).  The hybrid method
  produces a major improvement in $z\la0.5$ starburst galaxies, and,
  as quantified in Table \ref{Tab: stats_GAL}, reduces the overall
  dispersion by over 20\% compared with the standard template--fitting
  method.}
\label{Fig: hybrid_GAL}
\end{figure*}
Figure \ref{Fig: hybrid_GAL} compares the hybrid (top panel) and
template--fitting (bottom panel) photometric redshift predictions to
the spectroscopy for the AGES sample of galaxies (dots).  The
difference between the methods is subtle, but most evident at low
redshift ($z\sim 0.2$) where the improvement for starburst
PAH--emitters is apparent.  Also plotted is the in--house
spectroscopic galaxy sample (filled squares), which extends to much
fainter flux limits ($R\sim 25$) and higher redshifts than the AGES
galaxies.  As such it provides a valuable, independent test of the
method.  Clearly redshift estimation is reliable to $z\sim 1.5$ for
this deeper sample.  There are a handful of outliers between $0.5 \la
z \la 1$, and evidence of possible systematic errors at $z \la 0.25$
where the $B_W$ filter is not fully blueward of the 4000 \AA\ break.

Various measures of the accuracy of the hybrid and template--fitting
methods are given in Table \ref{Tab: stats_GAL}.  The basic result is
that a redshift accuracy of $\sigma \approx 0.06\,(1+z)$ is being
achieved for 99.5\% of galaxies in the AGES sample.  For the more
challenging in--house sample, 98.5\% of galaxies have a redshift
accuracy of $\sigma \approx 0.10\,(1+z)$.  The hybrid method also
reduces the number of outliers, as seen in the fraction of 3$\sigma$
outliers.  The final two columns of Table \ref{Tab: stats_GAL} report
the dispersions for subsamples constrained to contain 95\% of the
objects, thereby allowing a direct comparison of the hybrid and
template methods.  Clearly the hybrid method is superior, reducing the
dispersion by over 20\% compared with standard template-fitting.
\begin{deluxetable*}{lccccccccccc}
\tabletypesize{\normalsize}
\tablecaption{Photometric Redshift Accuracy for Galaxies\label{Tab: stats_GAL}}
\tablewidth{0pt}
\tablehead{
\colhead{} &\colhead{} && \multicolumn{2}{c}{Unclipped}
  && \multicolumn{3}{c}{3$\sigma$ Clipped} &&  \multicolumn{2}{c}{95\% Clipped} \\ \\[-0.2cm] 
\cline{4-5} \cline{7-9}   \cline{11-12} \\[-0.2cm] 
 \colhead{Sample} & \colhead{Algorithm} &&  \colhead{$\sigma$}
 &  \colhead{$\sigma/(1+z)$}   && \colhead{\% Rejected} & \colhead{$\sigma$}
 &  \colhead{$\sigma/(1+z)$}   &&  \colhead{$\sigma$}
 &  \colhead{$\sigma/(1+z)$}
}
\startdata
AGES             &  Hybrid && 0.143 & 0.105  && 0.52  & 0.077 & 0.060   && 0.060 & 0.047  \\
In--House        &  Hybrid && 0.397 & 0.185  && 1.46  & 0.160 & 0.096   && 0.101 & 0.059  \\
AGES + In--House &  Hybrid && 0.160 & 0.109  && 0.53  & 0.080 & 0.061   && 0.062 & 0.048  \\ 
\\
AGES             &  Template && 0.230 & 0.170  && 0.60  & 0.102 & 0.081 && 0.079 & 0.061  \\
In--House        &  Template && 0.498 & 0.253  && 2.51  & 0.190 & 0.111 && 0.127 & 0.081  \\
AGES + In--House &  Template && 0.245 & 0.174  && 0.69  & 0.104 & 0.082 && 0.081 & 0.062  \\
\enddata
\end{deluxetable*}
\begin{figure*}[bthp]
\plotone{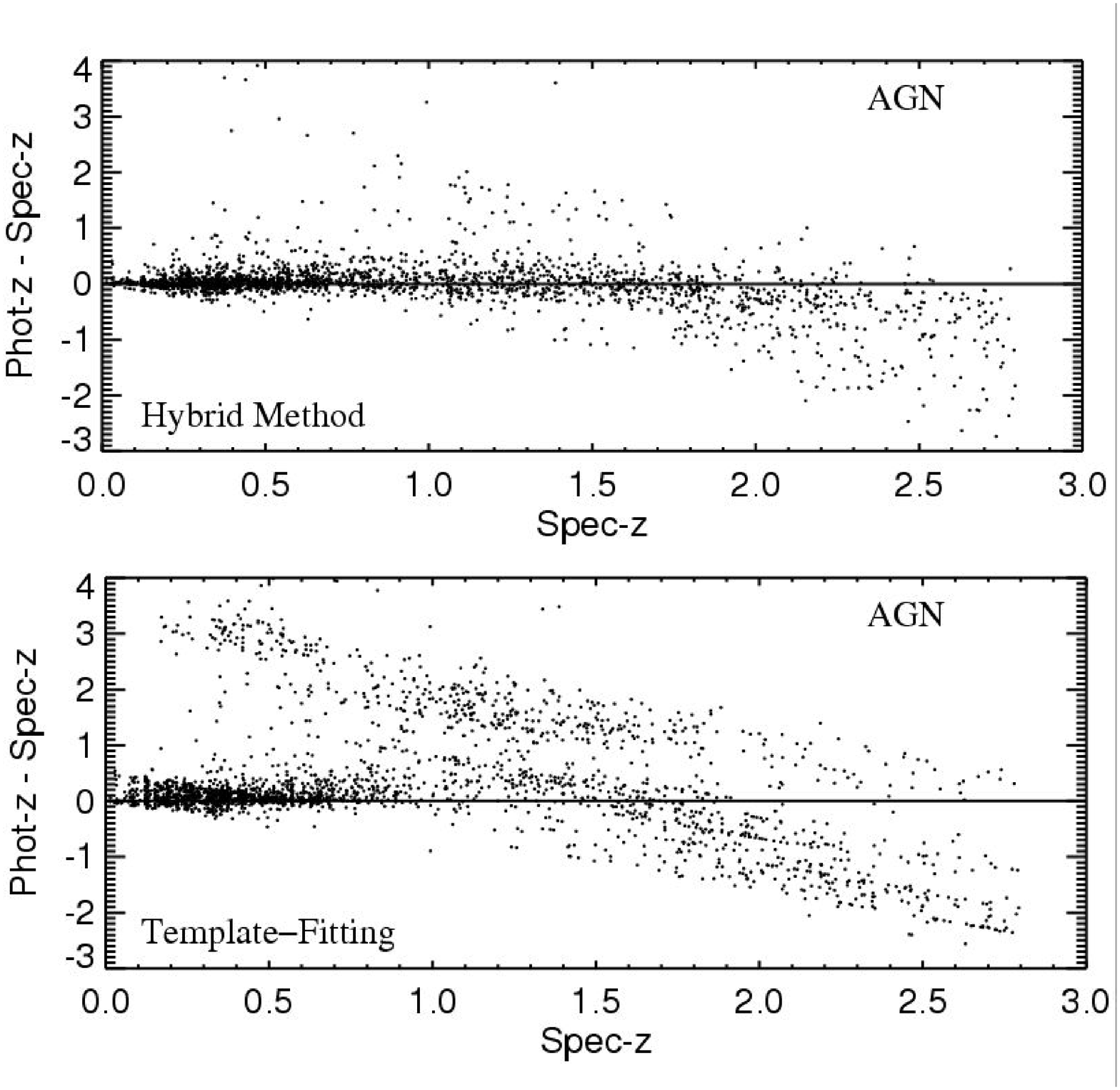}
\caption{Hybrid (top panel) and template--fitting (bottom panel)
  photometric redshifts for the AGN sample.  The improvement of the
  hybrid method is clear at all redshifts, dramatically so at $z>1$.
  A statistical comparison of the two methods, given in Table
  \ref{Tab: stats_AGN}, demonstrates that the hybrid method reduces
  the redshift dispersion by 65\% over the template--fitting method.}
\label{Fig: hybrid_AGN}
\end{figure*}
Figure \ref{Fig: hybrid_AGN} shows a similar comparison for the AGN
sample.  Here the improvement of the hybrid technique over simple
template--fitting is dramatic, though not surprising.  With no AGN
template in the mix, the template--fitting algorithm should not be
expected to succeed for accretion--dominated objects.  This is clear
from the lower panel of this figure, where beyond $z\ga1$ the results
are essentially useless.  The fairly accurate results at low redshift
are likely for galaxies which, though active, have luminosities
dominated by fusion processes.  In the bright AGES sample, these
galaxies would only be visible at modest redshifts, whereas the $z>1$
sample should be almost entirely composed of extremely luminous
quasars (some of which are also present at low redshift).

\begin{figure}[bthp]
\hspace*{-0.65cm}
\epsscale{1.25}
\plotone{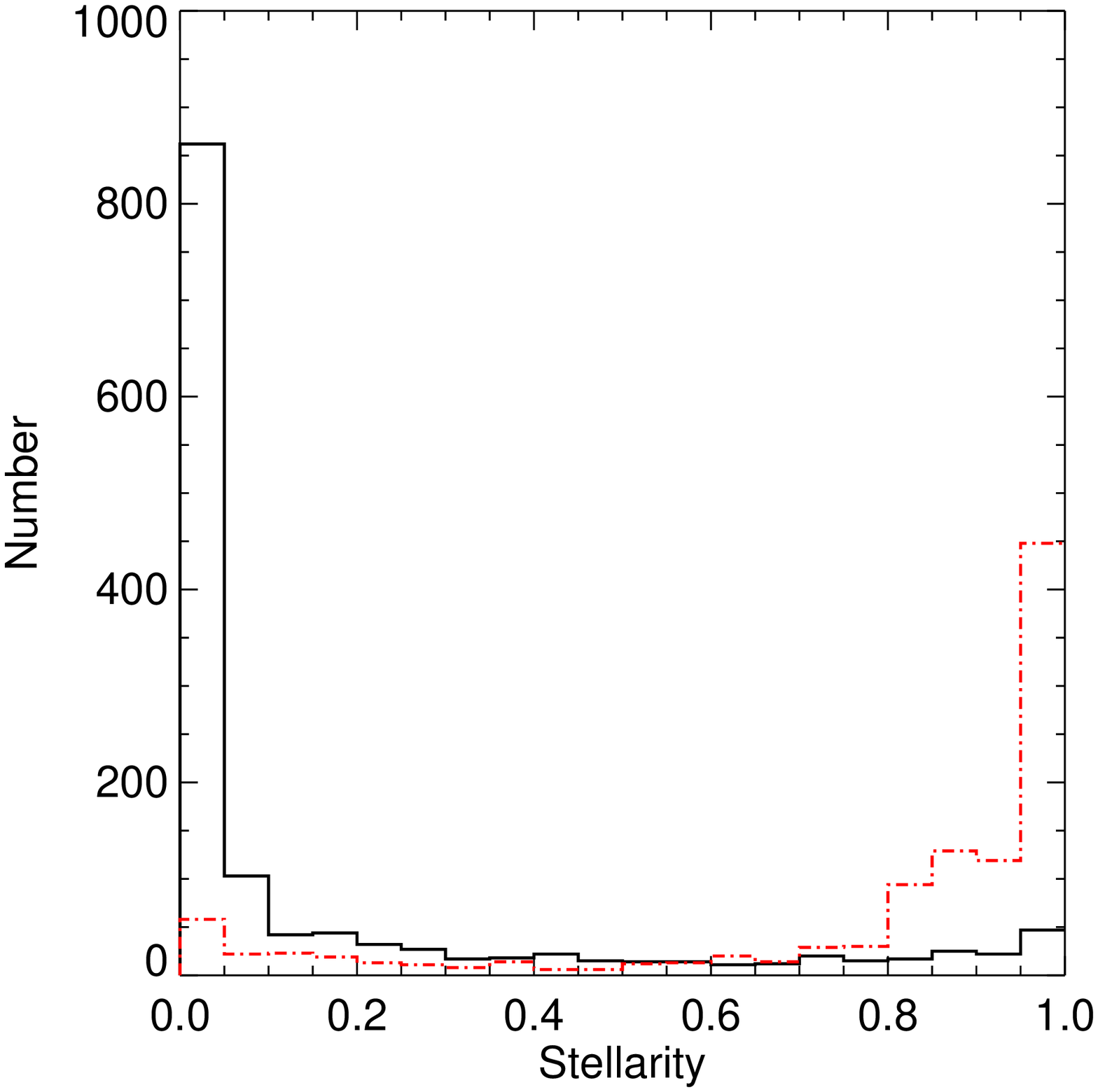}
\caption{Histogram of the SExtractor CLASS\_STAR parameter in the
  best-seeing optical band of two subsets of AGN isolated from the
  lower panel of Figure \ref{Fig: hybrid_AGN}.  This parameter varies
  continuously between zero and unity, where the extreme values
  correspond to sources which are completely resolved and unresolved,
  respectively.  The solid histogram represents those objects for
  which the template--fitting algorithm is reasonably robust, taken to
  be those objects with $|\mbox{Phot--}z - \mbox{Spec--}z| < 1$ and
  $\mbox{Spec--}z < 1$.  The dot--dashed histogram is for those
  objects excluded by this selection.  Clearly the template-fitting
  algorithm works well for resolved AGN and fails for truly
  quasi--stellar objects.}
\label{Fig: AGN Stellarity}
\end{figure}

As a check on this hypothesis, histograms of the morphological
stellarity indicator discussed above are plotted in Figure \ref{Fig:
  AGN Stellarity} for two AGN subsets, split according to their
observed redshift accuracy and range.  Those AGN for which the
template--fitting algorithm works reasonably well, taken to be those
with both spectroscopic redshifts and absolute redshift differences of
less than unity, are plotted as the solid histogram in Figure
\ref{Fig: AGN Stellarity}. The complement, for which the algorithm
largely fails, is plotted as a dot--dashed histogram.  Thus the
bimodality observed in the lower panel of Figure \ref{Fig: hybrid_AGN}
is strongly reproduced in the morphological stellarity measurements in
Figure \ref{Fig: AGN Stellarity}.  The AGN for which the
template--fitting method works are clearly resolved objects for which
the nuclear emission does not dominate the flux of the galaxy.
Conversely, the redshift failures are overwhelmingly unresolved
sources, consistent with the expectation of nuclear--dominated
emission from very luminous AGN and QSOs.
In marked contrast with the template-fitting redshifts, the hybrid
redshifts (top panel of Figure \ref{Fig: hybrid_AGN}) are quite
accurate for all AGN to $z\sim2$, with no obvious systematic issues or
significant occurrence of catastrophic errors.  Beyond $z>2$ the
redshifts for AGN are systematically underestimated, presumably due
the relative paucity of calibrators at these high redshifts.

Statistics of the redshift accuracy for AGN are given in Table
\ref{Tab: stats_AGN}.  The hybrid redshift dispersion is $\sigma
\approx 0.14\,(1+z)$ for over 97\% of the active galaxies.  The
tremendous improvement achieved by this method, apparent in Figure
\ref{Fig: hybrid_AGN}, is borne out in a direct statistical comparison
of the 95\%--clipped samples.  The dispersion of the hybrid method is
a factor of 3 smaller than that achieved with galaxy template fitting.
\begin{deluxetable*}{lccccccccccc}
\tabletypesize{\normalsize}
\tablecaption{Photometric Redshift Accuracy for AGN\label{Tab: stats_AGN}}
\tablewidth{0pt}
\tablehead{
\colhead{} &\colhead{} && \multicolumn{2}{c}{Unclipped}
  && \multicolumn{3}{c}{3$\sigma$ Clipped} &&  \multicolumn{2}{c}{95\% Clipped} \\ \\[-0.2cm] 
\cline{4-5} \cline{7-9}   \cline{11-12} \\[-0.2cm] 
 \colhead{Sample} & \colhead{Algorithm} &&  \colhead{$\sigma$}
 &  \colhead{$\sigma/(1+z)$}   && \colhead{\% Rejected} & \colhead{$\sigma$}
 &  \colhead{$\sigma/(1+z)$}   &&  \colhead{$\sigma$}
 &  \colhead{$\sigma/(1+z)$}
}
\startdata
AGES             &  Hybrid   && 0.473 & 0.219  && 2.94  & 0.309 & 0.138  && 0.255 & 0.120  \\
AGES             &  Template && 0.998 & 0.540  && 0.60  & 0.918 & 0.462  && 0.797 & 0.341  \\
\enddata
\end{deluxetable*}
\subsection{Comparison with an Independent Photometric Redshift Catalog}
\label{Sec: compare_brown}

We have also verified that the hybrid photometric redshift algorithm
presented here is in excellent agreement with an independent
photometric redshift catalog in the same field.  Brown \etal\ (in
preparation), adopting a pure neural net approach, have generated a
photometric redshift catalog from independently extracted multi--color
catalogs, photometered with an original code.  By making fainter
copies of the AGES spectroscopic galaxies, they have effectively
extended the calibration set to much fainter magnitudes.  In addition
to the photometric data, structural information, in the form of sizes
of the major and minor axes, was also incorporated into the neural net
for bright objects to improve accuracy at low redshift.  Using an
independent calibration of the neural net on this extended sample,
they derive neural net photometric redshifts for a large
optically-selected sample in Bo\"otes.

At relatively bright magnitudes ($R\le 23$) objects with SExtractor
CLASS\_STAR parameters greater than 0.85, measured in the best--seeing
optical band, are taken be stars (Jannuzi \etal\ in preparation) and
are removed for this comparison.  The stellar contamination is
negligible faintward of this limit and no attempt is made to remove
fainter stars.  Over the redshift range where spectroscopic
calibrators exist, $0 < z \la 1.5$, the inter--catalog 95\% clipped
redshift dispersion for the sample of galaxies common to both samples
is $\sigma = 0.09$, or $\sigma = 0.05\,(1+z)$.  These two sets of
redshifts were computed using independent galaxy photometry and error
estimates, and for the vast majority of objects, different photometric
redshift algorithms (template--fitting vs.\ artificial neural net).
Yet there is excellent agreement, similar in accuracy to that
demonstrated versus spectroscopy in the preceding section, probing
right down to the 13.3\muj\ limit.  This provides strong evidence that
both photometric redshift catalogs are free from substantial
systematic errors.

\subsection{Dependence on Magnitude and SED}

Comparing the results for the AGES and in-house galaxy samples in
Table \ref{Tab: stats_GAL}, it is clear that the redshift precision is
lower for the fainter in-house sample.  In general the photometric
redshift precision depends on both photometric S/N and galaxy spectral
type, with slightly smaller uncertainties typically achieved for
redder galaxies due to their larger continuum breaks.  Table \ref{Tab:
  s2n_sed} quantifies the hybrid redshift accuracy in differential
magnitude bins for galaxies classified in the template fits as earlier
or later than an unevolved CWW Sbc galaxy.
\begin{deluxetable}{ccccccccccc}
\tabletypesize{\normalsize}
\tablecaption{Dependence on Galaxy Magnitude and SED \label{Tab: s2n_sed}}
\tablewidth{0pt}
\tablehead{
\colhead{} && \multicolumn{2}{c}{Early--Type} &&  \multicolumn{2}{c}{Late--Type} \\ \\[-0.2cm] 
\cline{3-4} \cline{6-7} \\[-0.2cm] 
\colhead{4.5\mum\ Mag Range} &&  \colhead{$\sigma$}
 &  \colhead{$\sigma/(1+z)$}   &&  \colhead{$\sigma$}
 &  \colhead{$\sigma/(1+z)$}
}
\startdata
14.5 -- 15.0 && 0.039 & 0.031  && 0.038 & 0.030  \\ 
15.0 -- 15.5 && 0.038 & 0.029  && 0.046 & 0.034  \\ 
15.5 -- 16.0 && 0.040 & 0.029  && 0.049 & 0.036  \\ 
16.0 -- 16.5 && 0.050 & 0.037  && 0.071 & 0.054  \\ 
16.5 -- 17.0 && 0.069 & 0.054  && 0.083 & 0.063  \\ 
17.0 -- 17.5 && 0.074 & 0.060  && 0.073 & 0.057  \\ 

\enddata \tablecomments{The AGES + in--house galaxy sample is divided
  into early-- and late--type samples according to the best--fit SED
  templates; objects with best-fit templates earlier than an unevolved
  CWW Sbc template are classified as early--type, whereas those of
  type Sbc or later are taken to be late--type.  Statistics are for
  the 95\% clipped sample.}
\end{deluxetable}
The available spectroscopy beyond the AGES 4.5\mum\ statistical limit
of 15.7 mag is both incomplete and inhomogeneous.  The tabulated
redshift accuracies therefore may not be representative of those which
would be obtained for magnitude limited samples to fainter depths.

\newpage 
\section{Redshift Probability Functions}
\label{Sec: RPF}

Redshift likelihood functions were constructed by projecting the
likelihood surface of redshift and spectral type onto the redshift
axis.  Convolution of these likelihood functions with a
variable--width Gaussian kernel, which increases the error with
redshift, accounts for the template--mismatch variance inherent in the
method \citep[\eg][]{brodwin06,fernandez-soto02}.  From the results of
Table \ref{Tab: stats_GAL}, the kernel is taken to be $\sigma(z) =
0.06 (1+z)$.

This renders the area under these functions reasonable proxies of
redshift probability density, resulting in approximate redshift
probability distribution functions (PDFs) for {\it all} objects in the
survey.  Due to the excellent redshift accuracy for $0 < z < 1.5$, no
redshift prior was applied to these likelihood functions, other than
the weak prior imposed by the limited fitting range.  Stronger priors
can, of course, be imposed for certain kinds of analyses; we employ
this methodology in computing the galaxy redshift distribution in the
next section.  The statistical validity of the PDFs in the present
paper can be explicitly confirmed for those objects which have
spectroscopic redshifts as follows.

Redshift confidence intervals are derived from the PDFs by associating
area with probability density as illustrated in Figure \ref{Fig:
  Confidence}.  We define the 1$\sigma$, 2$\sigma$, and 3$\sigma$
confidence intervals as those redshift regions which enclose the top
68.3\%, 95.4\% and 99.7\% of the normalized area under the PDFs. 
\begin{figure*}[bthp]
\hspace*{-0.65cm}
\epsscale{1.25}
\plotone{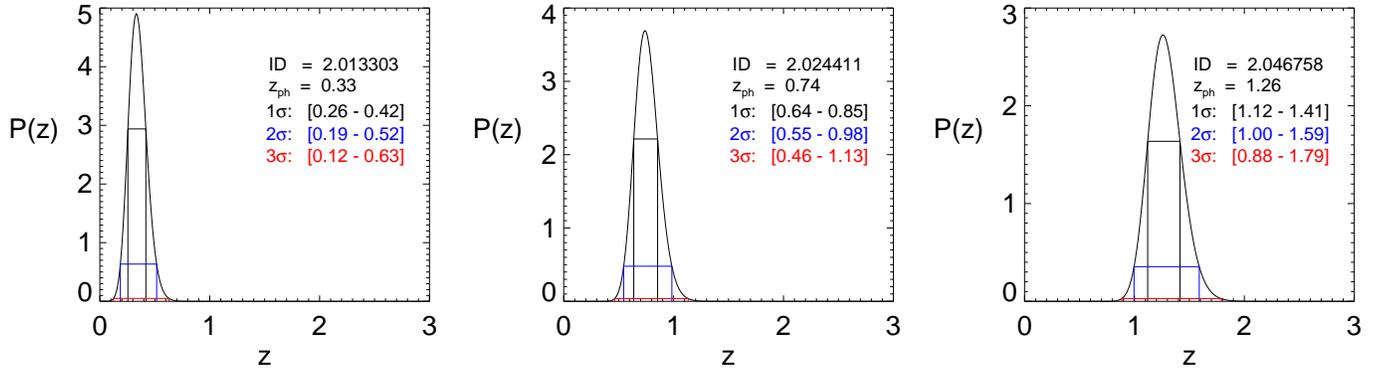}
\caption{Redshift PDFs for 3 sample galaxies, with the 1$\sigma$
  (68.3\%), 2$\sigma$ (95.4\%), and 3$\sigma$ (99.7\%) confidence
  intervals denoted by horizontal lines under the curves.}
\label{Fig: Confidence}
\end{figure*}
While the full PDF should in general be used for statistical analyses
of the galaxy population, confidence intervals defined in this way
offer a straightforward test of the PDFs.  In Table \ref{Tab: PDF
  test} we report the fraction of objects which agree with the
spectroscopic redshift at each confidence level, for both the full
spectroscopic sample and for the galaxy sample alone (\ie\ excluding
objects identified as QSOs and AGN in AGES, but including
PAH--emitting starburst galaxies).  We only consider objects here with
photometry of sufficient quality to be included in the main
photometric redshift sample as described in Section \ref{Sec: Photo-z
  Sample}.

The $1\sigma$ confidence intervals are approximately Gaussian for the
full sample in that the spectroscopic redshift is included in this
photometric redshift confidence interval for 70\% of the objects.  The
2$\sigma$ and 3$\sigma$ confidence intervals have inclusion rates
slightly below the Gaussian expectation, by 3.9\% and 2.7\%,
respectively.  This small outlier fraction is reduced to 0.8\% and
1.2\% when QSOs and AGN are excluded, though the $1\sigma$ interval is
slightly conservative in this case.  That the inclusion of a large
fraction of active galaxies affects the statistics so little is an
indication of the robustness of the redshift probability functions.
The lack of strong continuum spectral features results in quite broad
redshift probability functions for these objects, reflective of the
larger uncertainty in their redshifts.  These lead to wide, but valid,
confidence intervals, which can be incorporated in large statistical
studies.  On the other hand, previous IRAC Shallow survey papers
\citep{eisenhardt04,stern05} have demonstrated how these objects can
be reliably removed, if desired, using photometric information prior
to redshift fitting.
\begin{deluxetable*}{ccrccrc}
\tabletypesize{\small}
\tablecaption{Confidence Level Statistics\label{Tab: PDF test}}
\tablewidth{0pt}
\tablehead{
\colhead{}& \colhead{}& \multicolumn{2}{c}{All Objects}& \colhead{}& \multicolumn{2}{c}{Galaxies}\\
\cline{3-4} \cline{6-7} \\[-0.2cm] 
\colhead{Confidence} & \colhead{Gaussian}  &\colhead{Correct Within~~~} &
\colhead{Observed} &  \colhead{}& \colhead{Correct Within~~~} &
\colhead{Observed}  \\
\colhead{Level} & \colhead{Expectation} & \colhead{Confidence Interval} &
\colhead{Fraction} &\colhead{} & \colhead{Confidence Interval} &
\colhead{Fraction}}
\startdata
$\le 1\sigma$ &   68.3\% & 10870/15530 & 70.0\% &&  9722/13043 & 74.5\%\\
$\le 2\sigma$ &   95.4\% & 14206/15530 & 91.5\% && 12335/13043 & 94.6\%\\ 
$\le 3\sigma$ &   99.7\% & 15065/15530 & 97.0\% && 12848/13043 & 98.5\%\\
$> 3\sigma$   &    0.3\% &   465/15530 &  3.0\% &&   195/13043 & 1.5 \%
\enddata
\end{deluxetable*}
\section{Science Applications}
\label{Sec: Applications}

A shared primary goal of the IRAC Shallow, NDWFS and FLAMEX surveys is
to study structure formation and evolution at $0.5 < z < 2$, both in
the field and in cluster environments.  In this Section we present two
specific large scale structure science applications enabled by the
hybrid photometric redshifts and redshift probability distributions
derived in this paper: a measurement of the 4.5\mum\ galaxy redshift
distribution, and the discovery of a high--redshift ($z>1$) galaxy
cluster.

\subsection{Redshift Distribution at 4.5\mum}
\label{Sec: N(z)}
A key issue in structure formation models is the mass assembly history
of massive galaxies \citep[\eg][]{faber06}.  Recent work
\citep{yan05,mobasher05,bunker06} has indicated that massive galaxies
form the bulk of their stars soon after reionization, and evolve
passively over most of the history of the universe
\citep{treu05,juneau05,bundy05}.  While hierarchical models can
accommodate a modest number of early, massive halos, these results
were not predicted in advance of the observations.

More generally, any viable galaxy formation theory must predict the
low order moments of the mass distribution, including the redshift
distribution and the autocorrelation function.  Previous attempts to
constrain galaxy formation models using these moments have been
limited by the difficulty in relating optical light to mass, the
natural theoretical variable.  \citet{kauffmann98} suggested employing
a $K$--band selection to minimize this source of uncertainty, and
\citet[hereafter K20]{cimatti02} subsequently presented the $K_s < 20$
redshift distribution.  Their sample had a median redshift of $\zmed
\sim0.80$, and a distribution that was best reproduced by pure
luminosity evolution models.

\begin{figure*}[bthp]
\hspace*{-0.65cm}
\epsscale{0.8}
\plotone{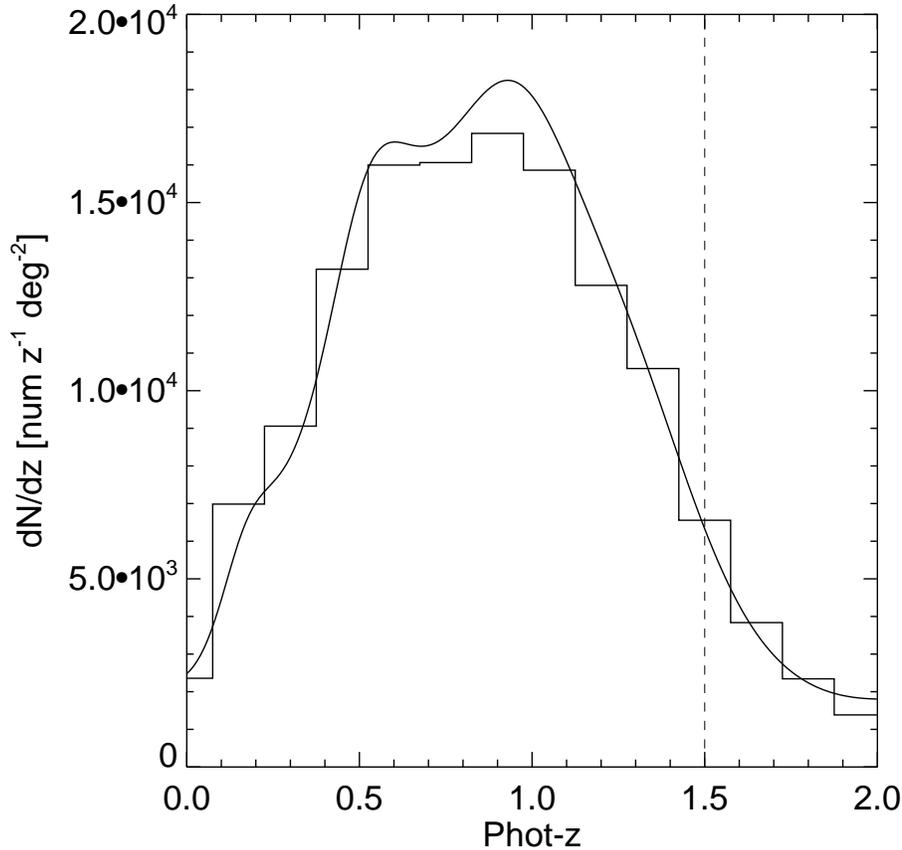}
\caption{Differential 4.5\mum--selected photometric redshift
  distribution in Bo\"otes.  The hybrid redshift catalog was used for
  the histogram.  The smooth curve was constructed from summing up the
  individual galaxy redshift probability functions as described in the
  text.  The dashed vertical line indicates the maximum redshift at
  which the photometric redshift method has been verified against
  spectroscopy.}
\label{Fig: N(z)}
\end{figure*}

However, the $K$--band is only a good proxy for stellar mass at
redshifts where it samples the rest--frame near-IR stellar peak.  At
$1<z<2$, where significant elliptical galaxy assembly is expected in
hierarchical models \citep[\eg][]{kauffmann96}, the stellar peak is
firmly in the \spitzer/IRAC bands.  We therefore present in Figure
\ref{Fig: N(z)} the 4.5\mum\ galaxy redshift distribution derived from
the 13.3\muj\ sample described above.  The survey is over 85\%
complete to this limit, based on the recovery fraction of artificial
stars in standard completeness simulations.  Very strict multi--band
masking was employed in determining the redshift distribution,
rejecting areas not containing valid coverage in {\em all} of the key
optical ($B_WRI$) and IRAC ($[3.6][4.5]$) bands.  This resulted in a
final unmasked area of 7.25 deg$^2$.  Stars are rejected at bright
magnitudes only, using the criteria described in
\textsection{\ref{Sec: compare_brown}}.
The redshift distribution is calculated in two ways.  The simple
histogram is derived using the best-estimate redshifts of the hybrid
method.  The curve in the figure shows the result of summing up the
full normalized redshift probability function for each galaxy,
following the method of \citet{brodwin06}.  The prior in this method
is taken to be the approximate $N(z)$ estimated from a direct
summation of the galaxy likelihood functions.  The vertical dashed
line represents the redshift limit of the available spectroscopy;
beyond this limit the robustness of the photometric redshifts has not
been explicitly demonstrated.  The median redshift is $\zmed = 0.98$
for the simple hybrid--method histogram, and $\zmed = 0.99$ for the
PDF summation method.  Assuming the galaxies with formal photometric
redshifts of $z\ge2$ are distributed according to the plotted $0<z<2$
distribution, the median redshifts are $\zmed = 0.85$ and $\zmed =
0.88$, respectively.  The two methods are in excellent agreement,
which, while not tautological, is nevertheless expected.  It provides
a measure of confidence in the consistency of the methods employed
here.

Despite being shallower than the K20 survey, the IRAC Shallow survey
has a higher median redshift, owing to a beneficial, negative
K--correction (\eg\ Eisenhardt et al. in preparation).  This deeper
reach, along with a 500--fold increase in area over the K20 survey,
enables stronger constraints to be placed on models of galaxy
evolution.  We will be examining this in detail in a future paper.

\subsection{A High Redshift Galaxy Cluster Search}
\label{Sec: cluster}

We present some early results of a high redshift ($z>1$) galaxy
cluster search underway in the Bo\"otes field.  The detection
technique, described in Eisenhardt \etal\ (in preparation, see also
\citealt{stanford05,elston06}; Gonzalez \etal\ in preparation),
implements a wavelet search algorithm tuned to identify structure on
cluster scales ($\sim 500$ kpc).  The redshift probability functions
are the input to the wavelet code, and so the cluster detection is
principally dependent on the accuracy and statistical reliability of
the redshift PDFs.  The method is independent of the strength, or even
{\it presence}, of the cluster red sequence, and therefore provides an
unbiased window on the era of cluster formation.

Four cluster candidates were targeted spectroscopically in the first
half of 2005, and all four were confirmed to be $z>1$ galaxy clusters,
at redshifts ranging from $\left<z\right>=1.11$ to
$\left<z\right>=1.41$.  The latter cluster, the highest yet found in a
cluster survey, is presented in \citet{stanford05}.  In this Section
we present one of the newly discovered clusters, a filamentary cluster
at $\left<z\right> = 1.24.$

\begin{figure*}[bthp]
\plotone{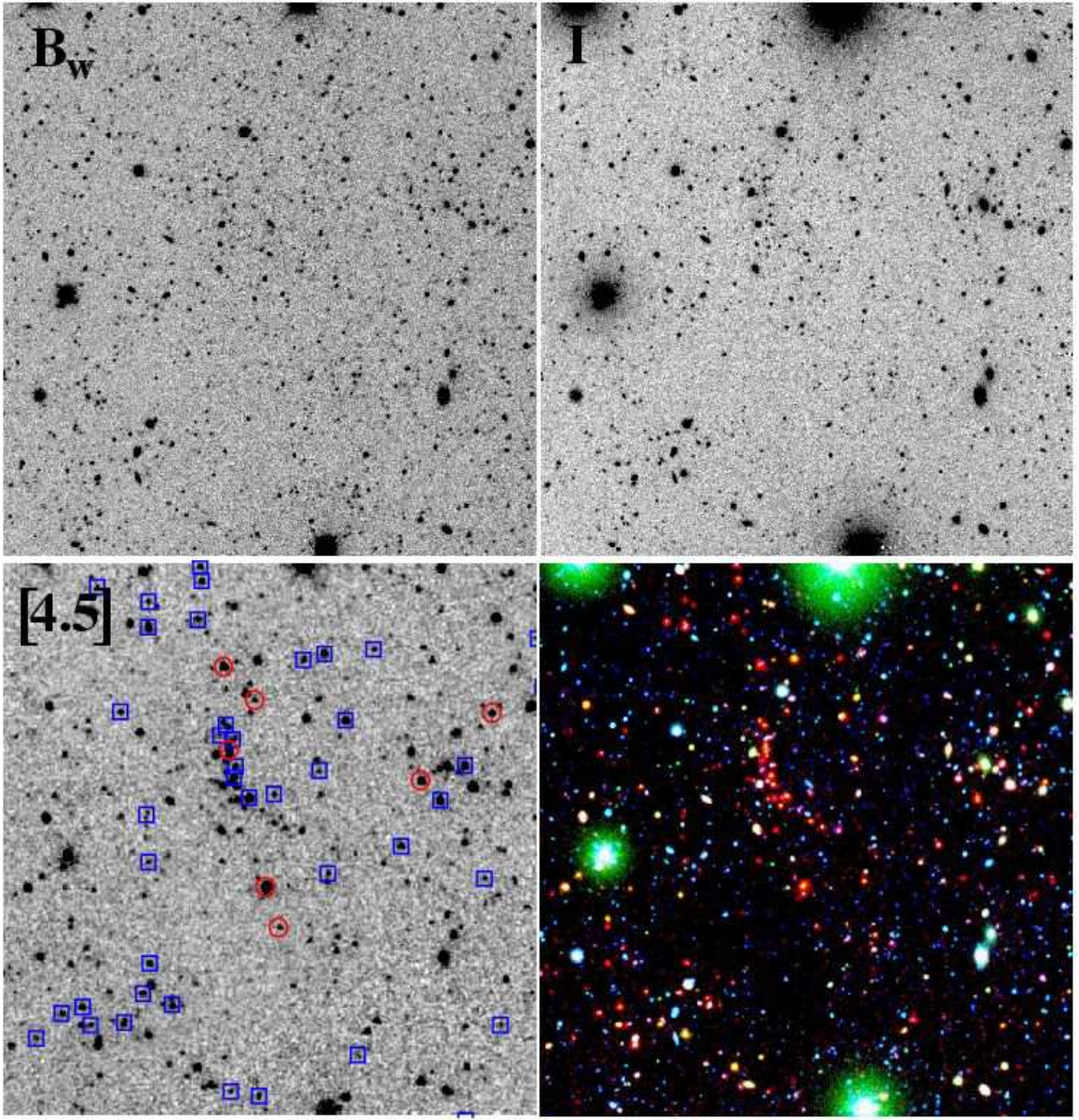}
\caption{5\arcmin$\times$5\arcmin\ $B_W$, $I$, $[4.5]$, and composite
  color image of cluster ISCS J1434.5+3427 at $\left<z\right>=1.24$.
  North is up and East is left.  Notice the striking filamentary
  morphology that emerges in the 4.5\mum\ band, and is quite prominent
  in the composite image.  Spectroscopically confirmed cluster members
  are denoted by red circles on the $[4.5]$ image.  Objects which
  contain the cluster systemic redshift within their 1$\sigma$
  photometric redshift confidence levels are denoted with blue
  squares.}
\label{Fig: Cluster}
\end{figure*}

The cluster ISCS J1434.5+3427 is shown in Figure \ref{Fig: Cluster}.
The cluster--finding power of the \spitzer/IRAC imaging is clearly
demonstrated in this figure.  Though no structure is apparent in the
optical bands, a striking filamentary structure emerges beyond
rest--frame 4000 \AA\ in the 4.5\mum\ band.  The spectroscopic cluster
members are circled in red on the greyscale 4.5\mum\ image.  Objects
which have the cluster systemic redshift of $\left<z\right> = 1.24$
within their 1$\sigma$ confidence intervals are marked by blue
squares.

ISCS J1434.5+3427 was observed spectroscopically in 2005 February with
Keck/LRIS \citep{lris} and 2005 May with Keck/DEIMOS \citep{deimos}.
For a more detailed description of these spectroscopic observation see
\citet{stanford05} and Desai \etal\ (in preparation).  The eight
spectroscopic members within $\Delta z = 0.01$ of the systemic
redshift confirm the reality of this cluster at
$\left<z\right>=1.241$.  Details of these members are given in Table
\ref{Tab: Spectroscopy}.  An estimate of the cluster velocity
dispersion is deferred until additional spectroscopy yields more
members.  In addition, deep follow--up imaging observations with {\it
  HST}/ACS and \spitzer/IRAC are underway.
\begin{deluxetable*}{lccrlcllcc}
\tabletypesize{\scriptsize}
\tablecaption{Summary of Spectroscopic Members\label{Tab: Spectroscopy}}
\tablewidth{0pt}
\tablehead{
&&&&&\multicolumn{1}{c}{95\% Confidence} &&&&\\
\multicolumn{1}{c}{ID} & \colhead{R.A.\tablenotemark{a}} &
\colhead{Dec.\tablenotemark{a}} & \multicolumn{1}{c}{$[4.5]$\tablenotemark{b}}&
\colhead{phot--$z$} & \multicolumn{1}{c}{Interval}&
\colhead{spec--$z$} & \multicolumn{1}{c}{$\delta z$}&  \colhead{Date}& \colhead{Instrument}
}
\startdata
IRAC J143430.3+342712 &  14:34:30.36  &  +34:27:12.1  & 15.81 & 1.26  & [0.98, 1.63]  & 1.2365 & 0.0005 & UT 2005 Feb 10 & LRIS\\
IRAC J143428.6+342557 &  14:34:28.66  &  +34:25:57.7  & 15.19 & 1.12  & [0.89, 1.43]  & 1.238  & 0.003  & UT 2005 Feb 10 & LRIS\\
IRAC J143421.6+342656 &  14:34:21.64  &  +34:26:56.2  & 16.67 & 1.09  & [0.84, 1.37]  & 1.2502 & 0.0005 & UT 2005 Feb 11 & LRIS\\
IRAC J143429.2+342739 &  14:34:29.21  &  +34:27:39.1  & 16.99 & 1.06  & [0.68, 1.49]  & 1.2436 & 0.0005 & UT 2005 Feb 11 & LRIS\\
IRAC J143430.1+342657 &  14:34:30.15  &  +34:26:57.2  & 15.90 & 1.17  & [0.92, 1.48]  & 1.23   & 0.01   & UT 2005 Feb 11 & LRIS\\
IRAC J143428.0+342535 &  14:34:28.05  &  +34:25:35.8  & 16.64 & 1.19  & [0.89, 1.57]  & 1.240\tablenotemark{c}  & 0.002  & UT 2005 May 07 & DEIMOS\\ 
IRAC J143418.4+342733 &  14:34:18.41  &  +34:27:33.1  & 16.42 & 1.21  & [0.96, 1.52]  & 1.240  & 0.001  & UT 2005 May 07 & DEIMOS\\
IRAC J143430.6+342757\tablenotemark{d} &  14:34:30.63  &  +34:27:57.2  & 15.35 & 0.95\tablenotemark{e}   & ~~[0.00, 4.17]\tablenotemark{c}    & 1.242  & 0.002 & UT 2005 May 07 & DEIMOS\\
\enddata
\tablenotetext{a}{Coordinates are J2000.}
\tablenotetext{b}{Vega magnitude at 4.5\mum; 0 mag = 179.5 Jy.}
\tablenotetext{c}{Lower quality redshift due to sky lines superimposed on \OII\ feature.}
\tablenotetext{d}{MIPS source.}
\tablenotetext{e}{Photometric redshift for this object is from the neural network.}
\end{deluxetable*}
\section{Summary}
\label{Sec: Summary}

Accurate photometric redshifts, calibrated using over 15,000
spectroscopic redshifts, have been computed for a 4.5\mum\ sample of
194,466 galaxies in the 8.5 deg$^2$ IRAC Shallow survey.  A hybrid
technique, in which a standard template fitting code is augmented
using a neural net approach, was adopted to optimize the redshift
accuracy for both active and low--$z$ starburst galaxies without
compromising reliability for the general galaxy population.  This is
primarily enabled by the fact that these two populations, which are
often troublesome for template-fitting codes, are particularly well
represented in the AGES sample.  Between $0 < z < 1.5$ the resulting
hybrid algorithm has a demonstrated accuracy of $\sigma \la
0.06\,(1+z)$ for 95\%, and $\sigma \la 0.1\,(1+z)$ for 98.5\%, of the
galaxy population.  For over 97\% of the active galaxies, the redshift
accuracy is $\sigma \la 0.14\,(1+z)$.

Redshift probability functions have been computed for all objects
directly from the template--fitting algorithm.  Comparison with the
large spectroscopic sample has verified the statistical validity of
these functions, and in particular, the reliability of confidence
intervals derived from them.  These confidence intervals, or indeed
the full probability functions, can be reliably used in statistical
studies of the galaxy population.  Several such programs are underway, 
and we present in this paper two new results which employ them.

The 4.5\mum-selected galaxy redshift distribution, a primary
observable for confronting theories of structure formation, was
computed using both the hybrid photometric redshifts and the full
redshift probability functions.  The methods yield entirely consistent
results.  This measurement is provided in anticipation of future model
predictions extending into the mid--IR, where flux is closely related
to stellar mass above $z\ga1$.

Another program making extensive use of these redshift PDFs is a
search for high redshift ($z>1$) galaxy clusters.  We presented one
such cluster, ISCS J1434.5+3427, spectroscopically confirmed at
$\left<z\right>=1.24$, which was discovered by incorporating the
redshift PDFs in a wavelet search algorithm.  Spectroscopic
confirmations of two similarly discovered high redshift clusters, at
$\left<z\right>=1.11$ and $\left<z\right>=1.41$, are presented in
companion papers \citep{stanford05, elston06}.  A complete description
of the cluster survey sample and methodology is presented in
Eisenhardt \etal\ (in preparation), along with the spectroscopic
confirmation of a $\left<z\right>=1.37$ cluster.

\acknowledgments 
We are grateful and indebted to the teams who built
the instruments used in this project.  This work is based in part on
observations made with the {\it Spitzer Space Telescope}, which is
operated by the Jet Propulsion Laboratory, California Institute of
Technology under a contract with NASA.
This paper made use of data from the NOAO Deep Wide-Field Survey
(NDWFS), which was supported by NOAO, AURA, Inc., and the National
Science Foundation.  We thank the entire NDWFS survey team and those
that operate and maintain Kitt Peak National Observatory, whose
facilities were used to obtain the NDWFS Bo\"otes field images.
%
%
Spectroscopic redshifts that contributed to the training of our
photometric redshift algorithms were obtained by numerous groups.  In
addition to the vast number of redshifts from the AGES survey
described in the text, redshifts were contributed by various projects,
including ones led by Hyron Spinrad, Steve Dawson, Richard Green,
James Rhoads, Sangeeta Malhotra, and George Djorgovski. We are
grateful to them for generously sharing their redshifts with us.
We are grateful to the staff of the MMT, W.~M.~Keck, and KPNO
observatories for their help with obtaining the various spectroscopic
data.
AHG and EM acknowledge support from an NSF Small Grant for Exploratory
Research under award AST--0436681.  SAS's work was performed under the
auspices of the U.S. Department of Energy, National Nuclear Security
Administration by the University of California, Lawrence Livermore
National Laboratory under contract No. W-7405-Eng-48.

\bibliographystyle{astron2} \bibliography{bibfile}

\begin{thebibliography}{}

\bibitem[\protect\astroncite{{Babbedge} et~al.}{2004}]{babbedge04}
{Babbedge}, T.~S.~R., et al.  2004,
\newblock {\em \mnras,} {\bf 353}, 654

\bibitem[\protect\astroncite{{Ben{\'{\i}}tez}}{2000}]{benitez}
{Ben{\'{\i}}tez}, N. 2000,
\newblock {\em \apj,} {\bf 536}, 571

\bibitem[\protect\astroncite{{Bertin} and {Arnouts}}{1996}]{sextractor}
{Bertin}, E. and {Arnouts}, S. 1996,
\newblock {\em \aaps,} {\bf 117}, 393

\bibitem[\protect\astroncite{{Blake} et~al.}{2006}]{blake06}
{Blake}, C., {Collister}, A., {Bridle}, S., and {Lahav}, O. 2006,
\newblock {\em ArXiv Astrophysics e-prints} \mnras submitted
  (astro--ph/0605303)

\bibitem[\protect\astroncite{{Brand} et~al.}{2006}]{xbootes3}
{Brand}, K., et al.  2006,
\newblock {\em \apj,} in press (astro--ph/0512343)

\bibitem[\protect\astroncite{{Brodwin} et~al.}{2006}]{brodwin06}
{Brodwin}, M., {Lilly}, S.~J., {Porciani}, C., {McCracken}, H.~J., {Le
  F{\`e}vre}, O., {Foucaud}, S., {Crampton}, D., and {Mellier}, Y. 2006,
\newblock {\em \apjs,} {\bf 162}, 20

\bibitem[\protect\astroncite{{Brunner} et~al.}{2000}]{brunner00}
{Brunner}, R.~J., {Szalay}, A.~S., and {Connolly}, A.~J. 2000,
\newblock {\em \apj,} {\bf 541}, 527

\bibitem[\protect\astroncite{{Bruzual} and {Charlot}}{2003}]{bc03}
{Bruzual}, G. and {Charlot}, S. 2003,
\newblock {\em \mnras,} {\bf 344}, 1000

\bibitem[\protect\astroncite{{Bundy} et~al.}{2005}]{bundy05}
{Bundy}, K., {Ellis}, R.~S., and {Conselice}, C.~J. 2005,
\newblock {\em \apj,} {\bf 625}, 621

\bibitem[\protect\astroncite{{Bunker} et~al.}{2006}]{bunker06}
{Bunker}, A., {Stanway}, E., {Ellis}, R., {McMahon}, R., {Eyles}, L., and
  {Lacy}, M. 2006,
\newblock {\em New Astronomy Review} {\bf 50}, 94

\bibitem[\protect\astroncite{{Cimatti} et~al.}{2002}]{cimatti02}
{Cimatti}, A., et al.  2002,
\newblock {\em \aap,} {\bf 391}, L1

\bibitem[\protect\astroncite{{Collister} and {Lahav}}{2004}]{ANNz}
{Collister}, A.~A. and {Lahav}, O. 2004,
\newblock {\em \pasp,} {\bf 116}, 345

\bibitem[\protect\astroncite{{Connolly} et~al.}{1995}]{connolly95}
{Connolly}, A.~J., {Csabai}, I., {Szalay}, A.~S., {Koo}, D.~C., {Kron}, R.~G.,
  and {Munn}, J.~A. 1995,
\newblock {\em \aj,} {\bf 110}, 2655

\bibitem[\protect\astroncite{{Connolly} et~al.}{1997}]{connolly97}
{Connolly}, A.~J., {Szalay}, A.~S., {Dickinson}, M., {Subbarao}, M.~U., and
  {Brunner}, R.~J. 1997,
\newblock {\em \apjl,} {\bf 486}, L11

\bibitem[\protect\astroncite{{Cool} et~al.}{2006}]{cool06}
{Cool}, R.~J., et al.  2006,
\newblock {\em \apj,} in press (astro--ph/0605030)

\bibitem[\protect\astroncite{{Devriendt} et~al.}{1999}]{devriendt99}
{Devriendt}, J.~E.~G., {Guiderdoni}, B., and {Sadat}, R. 1999,
\newblock {\em \aap,} {\bf 350}, 381

\bibitem[\protect\astroncite{{Eisenhardt} et~al.}{2004}]{eisenhardt04}
{Eisenhardt}, P.~R., et al.  2004,
\newblock {\em \apjs,} {\bf 154}, 48

\bibitem[\protect\astroncite{{Elston} et~al.}{2006}]{elston06}
{Elston}, R.~J., et al.  2006,
\newblock {\em \apj,} {\bf 639}, 816

\bibitem[\protect\astroncite{{Faber} et~al.}{2003}]{deimos}
{Faber}, S.~M., et al.  2003,
\newblock in {\em Instrument Design and Performance for Optical/Infrared
  Ground-based Telescopes. Edited by Iye, Masanori; Moorwood, Alan F. M.
  Proceedings of the SPIE, Volume 4841, pp. 1657-1669 (2003).}, pp 1657--1669

\bibitem[\protect\astroncite{{Faber} et~al.}{2006}]{faber06}
{Faber}, S.~M., et al.  2006,
\newblock {\em \apj,} submitted (astro--ph/0506044)

\bibitem[\protect\astroncite{{Fabricant} et~al.}{2005}]{hectospec}
{Fabricant}, D., et al.  2005,
\newblock {\em \pasp,} {\bf 117}, 1411

\bibitem[\protect\astroncite{{Fernandez-Soto} et~al.}{2002}]{fernandez-soto02}
{Fernandez-Soto}, A., {Lanzetta}, K.~M., {Chen}, H.-W., {Levine}, B., and
  {Yahata}, N. 2002,
\newblock {\em \mnras,} {\bf 330}, 889

\bibitem[\protect\astroncite{{Fioc} and {Rocca-Volmerange}}{1999}]{pegase2}
{Fioc}, M. and {Rocca-Volmerange}, B. 1999,
\newblock {\em ArXiv Astrophysics e-prints}

\bibitem[\protect\astroncite{{Firth} et~al.}{2003}]{firth03}
{Firth}, A.~E., {Lahav}, O., and {Somerville}, R.~S. 2003,
\newblock {\em \mnras,} {\bf 339}, 1195

\bibitem[\protect\astroncite{{Firth} et~al.}{2002}]{firth02}
{Firth}, A.~E., et al.  2002,
\newblock {\em \mnras,} {\bf 332}, 617

\bibitem[\protect\astroncite{{Fontana} et~al.}{2000}]{fontana00}
{Fontana}, A., {D'Odorico}, S., {Poli}, F., {Giallongo}, E., {Arnouts}, S.,
  {Cristiani}, S., {Moorwood}, A., and {Saracco}, P. 2000,
\newblock {\em \aj,} {\bf 120}, 220

\bibitem[\protect\astroncite{{Helou} et~al.}{2000}]{helou00}
{Helou}, G., {Lu}, N.~Y., {Werner}, M.~W., {Malhotra}, S., and {Silbermann}, N.
  2000,
\newblock {\em \apjl,} {\bf 532}, L21

\bibitem[\protect\astroncite{{Hsieh} et~al.}{2005}]{rcs_photo-z}
{Hsieh}, B.~C., {Yee}, H.~K.~C., {Lin}, H., and {Gladders}, M.~D. 2005,
\newblock {\em \apjs,} {\bf 158}, 161

\bibitem[\protect\astroncite{{Jannuzi} and {Dey}}{1999}]{ndwfs99}
{Jannuzi}, B.~T. and {Dey}, A. 1999,
\newblock in {\em ASP Conf. Ser. 191 --- Photometric Redshifts and the
  Detection of High Redshift Galaxies}, p. 111

\bibitem[\protect\astroncite{{Juneau} et~al.}{2005}]{juneau05}
{Juneau}, S., et al.  2005,
\newblock {\em \apjl,} {\bf 619}, L135

\bibitem[\protect\astroncite{{Kauffmann}}{1996}]{kauffmann96}
{Kauffmann}, G. 1996,
\newblock {\em \mnras,} {\bf 281}, 487

\bibitem[\protect\astroncite{{Kauffmann} and {Charlot}}{1998}]{kauffmann98}
{Kauffmann}, G. and {Charlot}, S. 1998,
\newblock {\em \mnras,} {\bf 297}, L23+

\bibitem[\protect\astroncite{{Kenter} et~al.}{2005}]{xbootes2}
{Kenter}, A., et al.  2005,
\newblock {\em \apjs,} {\bf 161}, 9

\bibitem[\protect\astroncite{{Kinney} et~al.}{1996}]{kinney96}
{Kinney}, A.~L., {Calzetti}, D., {Bohlin}, R.~C., {McQuade}, K.,
  {Storchi--Bergmann}, T., and {Schmitt}, H.~R. 1996,
\newblock {\em \apj,} {\bf 467}, 38

\bibitem[\protect\astroncite{{Kitsionas} et~al.}{2005}]{kitsionas05}
{Kitsionas}, S., {Hatziminaoglou}, E., {Georgakakis}, A., and
  {Georgantopoulos}, I. 2005,
\newblock {\em \aap,} {\bf 434}, 475

\bibitem[\protect\astroncite{{Lacy} et~al.}{2004}]{lacy04}
{Lacy}, M., et al.  2004,
\newblock {\em \apjs,} {\bf 154}, 166

\bibitem[\protect\astroncite{{Lahav} et~al.}{1996}]{lahav96}
{Lahav}, O., {Naim}, A., {Sodr{\' e}}, L., and {Storrie-Lombardi}, M.~C. 1996,
\newblock {\em \mnras,} {\bf 283}, 207

\bibitem[\protect\astroncite{{Lu} et~al.}{2003}]{lu03}
{Lu}, N., et al.  2003,
\newblock {\em \apj,} {\bf 588}, 199

\bibitem[\protect\astroncite{{Makovoz} and {Khan}}{2005}]{mopex}
{Makovoz}, D. and {Khan}, I. 2005,
\newblock in P. {Shopbell}, M. {Britton}, and R. {Ebert} (eds.), {\em
  Astronomical Society of the Pacific Conference Series}, p.~81

\bibitem[\protect\astroncite{{Maraston}}{2005}]{maraston05}
{Maraston}, C. 2005,
\newblock {\em \mnras,} {\bf 362}, 799

\bibitem[\protect\astroncite{{Mobasher} et~al.}{2005}]{mobasher05}
{Mobasher}, B., et al.  2005,
\newblock {\em \apj} {\bf 635}, 832

\bibitem[\protect\astroncite{{Monet} et~al.}{2003}]{monet03}
{Monet}, D.~G., et al.  2003,
\newblock {\em \aj,} {\bf 125}, 984

\bibitem[\protect\astroncite{{Murray} et~al.}{2005}]{xbootes1}
{Murray}, S.~S., et al.  2005,
\newblock {\em \apjs,} {\bf 161}, 1

\bibitem[\protect\astroncite{{Oke} et~al.}{1995}]{lris}
{Oke}, J.~B., et al.  1995,
\newblock {\em \pasp,} {\bf 107}, 375

\bibitem[\protect\astroncite{{Padmanabhan} et~al.}{2006}]{padmanabhan06}
{Padmanabhan}, N., et al.  2006,
\newblock {\em ArXiv Astrophysics e-prints} \mnras submitted
  (astro--ph/0605302)

\bibitem[\protect\astroncite{{Pahre} et~al.}{2004}]{pahre04}
{Pahre}, M.~A., {Ashby}, M.~L.~N., {Fazio}, G.~G., and {Willner}, S.~P. 2004,
\newblock {\em \apjs,} {\bf 154}, 235

\bibitem[\protect\astroncite{{Reach} et~al.}{2005}]{reach05}
{Reach}, W.~T., et al.  2005,
\newblock {\em \pasp,} {\bf 117}, 978

\bibitem[\protect\astroncite{{Sawicki}}{2002}]{sawicki02}
{Sawicki}, M. 2002,
\newblock {\em \aj,} {\bf 124}, 3050

\bibitem[\protect\astroncite{{Sawicki} et~al.}{1997}]{sawicki97}
{Sawicki}, M.~J., {Lin}, H., and {Yee}, H.~K.~C. 1997,
\newblock {\em \aj,} {\bf 113}, 1

\bibitem[\protect\astroncite{{Simpson} and {Eisenhardt}}{1999}]{simpson99}
{Simpson}, C. and {Eisenhardt}, P. 1999,
\newblock {\em \pasp,} {\bf 111}, 691

\bibitem[\protect\astroncite{{Stanford} et~al.}{2005}]{stanford05}
{Stanford}, S.~A., et al.  2005,
\newblock {\em \apjl,} {\bf 634}, L129

\bibitem[\protect\astroncite{{Stern} et~al.}{2005}]{stern05}
{Stern}, D., et al.  2005,
\newblock {\em \apj,} {\bf 631}, 163

\bibitem[\protect\astroncite{{Treu} et~al.}{2005}]{treu05}
{Treu}, T., {Ellis}, R.~S., {Liao}, T.~X., and {van Dokkum}, P.~G. 2005,
\newblock {\em \apjl,} {\bf 622}, L5

\bibitem[\protect\astroncite{{Vanzella} et~al.}{2004}]{vanzella04}
{Vanzella}, E., et al.  2004,
\newblock {\em \aap,} {\bf 423}, 761

\bibitem[\protect\astroncite{{Vazdekis} et~al.}{1996}]{vazdekis96}
{Vazdekis}, A., {Casuso}, E., {Peletier}, R.~F., and {Beckman}, J.~E. 1996,
\newblock {\em \apjs,} {\bf 106}, 307

\bibitem[\protect\astroncite{{V{\'a}zquez} and {Leitherer}}{2005}]{starburst99}
{V{\'a}zquez}, G.~A. and {Leitherer}, C. 2005,
\newblock {\em \apj,} {\bf 621}, 695

\bibitem[\protect\astroncite{{Weinstein} et~al.}{2004}]{weinstein04}
{Weinstein}, M.~A., et al.  2004,
\newblock {\em \apjs,} {\bf 155}, 243

\bibitem[\protect\astroncite{{Wolf} et~al.}{2003}]{combo17_LF}
{Wolf}, C., {Meisenheimer}, K., {Rix}, H.-W., {Borch}, A., {Dye}, S., and
  {Kleinheinrich}, M. 2003,
\newblock {\em \aap,} {\bf 401}, 73

\bibitem[\protect\astroncite{{Yan} et~al.}{2005}]{yan05}
{Yan}, H., et al.  2005,
\newblock {\em \apj,} {\bf 634}, 109

\end{thebibliography}
\end{document}